\documentclass[%
 reprint,
 amsmath,amssymb,
 aps,nofootinbib,
]{revtex4-1}
\usepackage{amsmath,amssymb,mathtools}
\usepackage[pdftex]{graphicx}
\usepackage[pdftex]{hyperref}
\usepackage[utf8x]{inputenc}
\usepackage[normalem]{ulem}
\usepackage{slashed}
\usepackage{color}
\usepackage{xcolor,cancel}
\usepackage{array}
\usepackage{braket}
\usepackage{dcolumn}% Align table columns
\usepackage{caption}
\usepackage{subcaption}
\usepackage{bigstrut}
\usepackage{multirow}
\usepackage{dcolumn}% Align table columns on decimal point
\usepackage{bm}% bold math
\def\tev{\, {\rm TeV}}
\def\gev{\, {\rm GeV}}
\def\fb{\, {\rm fb}}
\def\lq{\rm LQ}
\def\bq{\rm b}
\def\tq{\rm t}
\def\wb{\rm W}
\def\zb{\rm Z}
\def\pt{p_{\rm{T}}}
\def\el{\rm e}
\def\Bm{\rm B}

\usepackage{color}

\begin{document}

\preprint{APS/123-QED}

\title{On the sensitivity reach of $\lq$ production with preferential couplings to third generation fermions at the LHC }% Force line breaks with \\

\author{A. Flórez}%
\email{ca.florez@uniandes.edu.co}
\affiliation{Departamento de F\'isica, Universidad de Los Andes, Cra. 1 \# 18a-12, Bogot\'a, Colombia}
\author{J. Jones-P\'erez}%
\email{jones.j@pucp.edu.pe}
\affiliation{Secci\'on F\'isica, Departamento de Ciencias, Pontificia Universidad Cat\'olica del Per\'u, Apartado 1761, Lima, Peru}
\author{A. Gurrola}%
\email{alfredo.gurrola@vanderbilt.edu}
\affiliation{Department of Physics and Astronomy, Vanderbilt University, Nashville, TN, 37235, USA}
\author{C. Rodriguez}%
\email{c.rodriguez45@uniandes.edu.co}
\affiliation{Departamento de F\'isica, Universidad de Los Andes, Cra. 1 \# 18a-12, Bogot\'a, Colombia}
\author{J. Peñuela-Parra}%
\email{j.penuela@uniandes.edu.co}
\affiliation{Departamento de F\'isica, Universidad de Los Andes, Cra. 1 \# 18a-12, Bogot\'a, Colombia}

%\affiliation{%
% Universidad de los Andes\\

%\collaboration{CMS-Colaboration}%\noaffiliation

%\affiliation{
%\affiliation{
%\affiliation{%


\date{\today}% It is always \today, today,

\begin{abstract}

Leptoquarks ($\lq$s) are hypothetical particles that appear in various extensions of the Standard Model (SM), that can explain observed differences between SM theory predictions and experimental results. The production of these particles has been widely studied at various experiments, most recently at the Large Hadron Collider (LHC), and stringent bounds have been placed on their masses and couplings, assuming the simplest beyond-SM (BSM) hypotheses. However, the limits are significantly weaker for $\lq$ models with family non-universal couplings containing enhanced couplings to third-generation fermions. We present a new study on the production of a $\lq$ at the LHC, with preferential couplings to third-generation fermions, considering proton–proton collisions at $\sqrt{s} = 13 \tev$ and $\sqrt{s} = 13.6 \tev$. Such a hypothesis is well motivated theoretically and it can explain the recent anomalies in the precision measurements of $\Bm$-meson decay rates, specifically the $R_{D^{(*)}}$ ratios. Under a simplified model where the $\lq$ masses and couplings are free parameters, we focus on cases where the $\lq$ decays to a $\tau$ lepton and a $\bq$ quark, and study how the results are affected by different assumptions about chiral currents and interference effects with other BSM processes with the same final states, such as diagrams with a heavy vector boson, $\zb^{\prime}$. The analysis is performed using machine learning techniques, resulting in an increased discovery reach at the LHC, allowing us  to probe  new physics phase space which addresses the $\Bm$-meson anomalies, for $\lq$ masses up to $5.00\tev$, for the high luminosity LHC scenario.

\end{abstract}

\maketitle

%\tableofcontents

\section{\label{sec:level1}Introduction}
\label{sec:intro}

After more than ten years collecting data, the LHC has confirmed that the Standard Model (SM) is indeed the correct theory describing particle physics for energies below the $\tev$ scale. Nevertheless, there exist reasons to expect the SM to be a low-energy effective realization of a more complete theory. On the theoretical side, we do not know if gravity should be quantized, or if the gauge interactions should be unified, and if so, we do not know how to solve the associated hierarchy problems on the Higgs mass. Moreover, we have no explanation for fermion family replication, nor for the lack of CP violation in the strong sector. This expectation for physics beyond the SM (BSM) is reinforced experimentally, where the observation of neutrino masses, dark matter, and the baryon asymmetry in the Universe, cannot be explained by the SM.

Leptoquarks ($\lq$s) are hypothetical bosons carrying both baryon and lepton number, thus interacting jointly with a lepton and a quark. They are a common ingredient in SM extensions where quarks and leptons share the same multiplet. Typical examples of these can be found in the Pati-Salam~\cite{Pati:1974yy} and $SU(5)$ GUT~\cite{Georgi:1974sy} models. In addition, they can also be found in theories with strong interactions, such as compositeness~\cite{Schrempp:1984nj}. Due to their exotic coupling which allows quark-lepton transitions, they have a diverse phenomenology, which naturally leads to several constraints. An important one comes from proton decay, which forces the $\lq$ mass to values close to the Planck scale, unless baryon and lepton numbers are not violated. Furthermore, in models where the latter are conserved, the $\lq$ can still be subject to a wide variety of bounds~\cite{Leurer:1993em,Davidson:1993qk,Leurer:1993qx,Hewett:1997ce,Queiroz:2014pra,Dorsner:2016wpm}. Examples of these come from meson mixing, electric and magnetic dipole moments, atomic parity violation tests, rare decays, and direct searches. Nevertheless, the significance of each bound is a model dependent question.

In the last years, an increased interest in low scale $\lq$s has emerged due to the anomalies in the precision measurements of the $\Bm$-meson decay rates. As it is well known, these corresponded mainly to deviations in the $R_{K^{(*)}}$~\cite{LHCb:2014vgu,LHCb:2017avl,LHCb:2019hip,LHCb:2021trn} and $R_{D^{(*)}}$~\cite{BaBar:2012obs,BaBar:2013mob,Abdesselam:2019dgh, Hirose:2017dxl, Sato:2016svk, Hirose:2016wfn, Huschle:2015rga,LHCb:2015gmp,Aaij:2015yra,Aaij:2017uff,LHCb:2017rln,LHCb:2023zxo} ratios, which measure the violation of lepton flavour universality (LFU). What followed was a very intense theoretical development, aiming to explain the anomalies by $\tev$ scale $\lq$ exchange at tree level~\cite{Hiller:2014yaa,Gripaios:2014tna,Alonso:2015sja,Calibbi:2015kma,Fajfer:2015ycq,Bauer:2015knc,Becirevic:2016oho,Crivellin:2017zlb,DAmico:2017mtc,Hiller:2017bzc,Buttazzo:2017ixm,Becirevic:2018afm,Cornella:2019hct,Angelescu:2021lln,Belanger:2021smw,GINO_2022}. Before the end of 2022, it was generally agreed that, within proposed single $\lq$ solutions, the only candidate capable of addressing all $\Bm$-meson anomalies simultaneously and surviving all other constraints was a vector $\lq$ ($U_1$), transforming as $({\bf 3},\,{\bf 1},\,2/3)$, and coupling mainly to third-generation fermions via $\bq\,\tau$ and $\tq\,\nu_\tau$ vertices~\cite{Buttazzo:2017ixm,Angelescu:2021lln}. In spite of a recent re-analysis of $R_{K^{(*)}}$ data showing this ratio to be compatible with the SM prediction~\cite{LHCb:2022qnv,LHCb:2022zom,Greljo:2022jac,Ciuchini:2022wbq}, the solution to the $R_{D^{(*)}}$ anomaly is still an open question and remains a valid motivation for the study of scenarios where new particles have preferential couplings to third-generation fermions. Thus, it is still of interest to continue exploring the possibility of observing the $U_1$ $\lq$ at the LHC~\cite{GINO_2022}. 

As expected, the theoretical community has extensively participated in probing $\lq$ models by scrutinizing search strategies, recasting LHC results, and predicting the reach in the parameter space via different searches involving third-generation fermions (see for instance~\cite{Diaz:2017lit,Dorsner:2018ynv,PhysRevD.99.035021,Schmaltz:2018nls,Biswas:2018snp,Baker:2019sli,Haisch:2020xjd,Bhaskar:2021gsy,Bernigaud:2021fwn,CompositenessGurrola}). In addition, several $13 \tev$ searches for $\lq$s decaying into $\tq/\bq$ and $\tau/\nu$ final states have been performed by the CMS~\cite{CMS:2016fxb,CMS:2017xcw,CMS:2018svy,CMS:2018qqq,CMS:2018txo,CMS:2018iye,CMS:2020wzx,CMS:2022goy,LQS_CMS_2022_results_comparison} and ATLAS~\cite{ATLAS:2019qpq,ATLAS:2020dsf,ATLAS:2021oiz,ATLAS:2021yij,ATLAS:2021jyv,ATLAS_7A,ATLAS_Vertical_Line} collaborations.

%Leptoquark Feynman Diagrams - fig:feynmp-prod-channels
\begin{figure}[!t]
    \centering
    %Single Leptoquark Production Diagram
    \begin{subfigure}[b]{0.23\textwidth}
        \includegraphics[width = 1.1\textwidth]{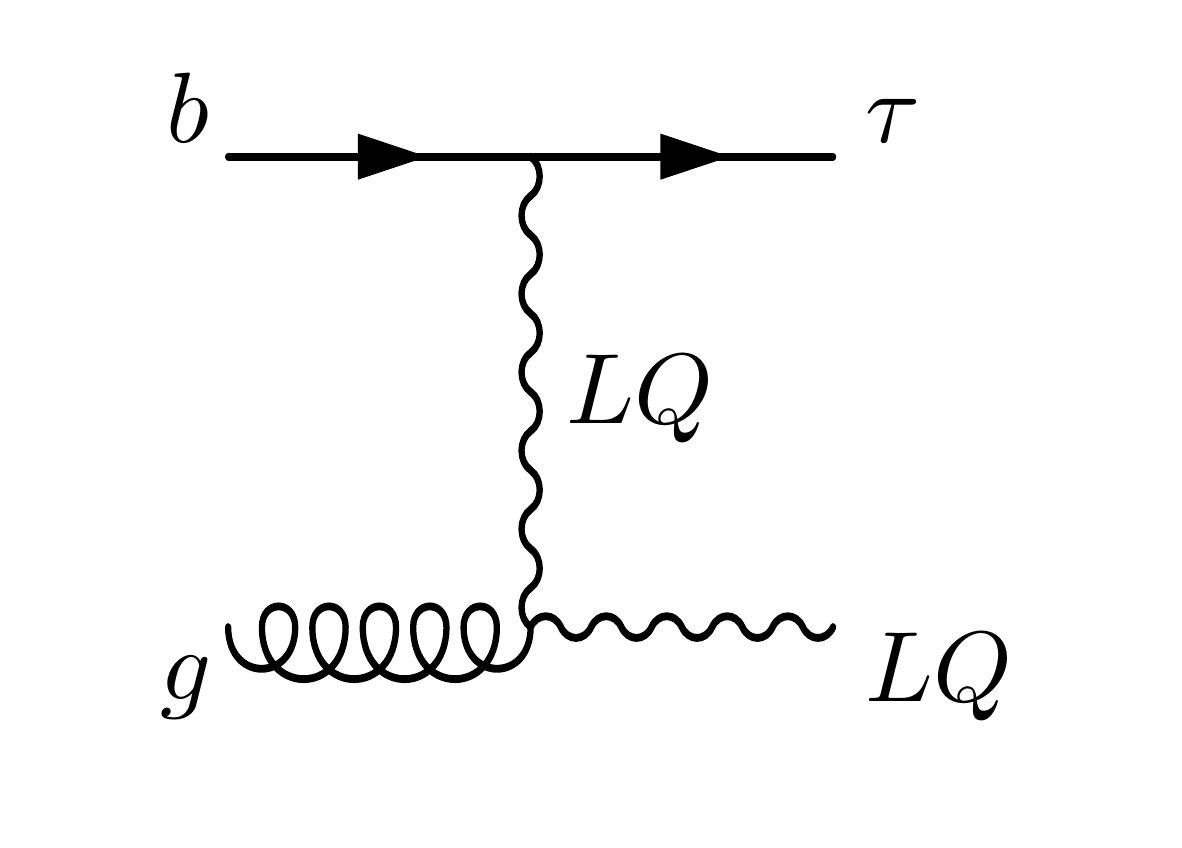}
        \caption{}
    \end{subfigure}
    \hfill
    %Double Leptoquark Production Diagram
    \begin{subfigure}[b]{0.23\textwidth}
        \includegraphics[width =  1.1\textwidth]{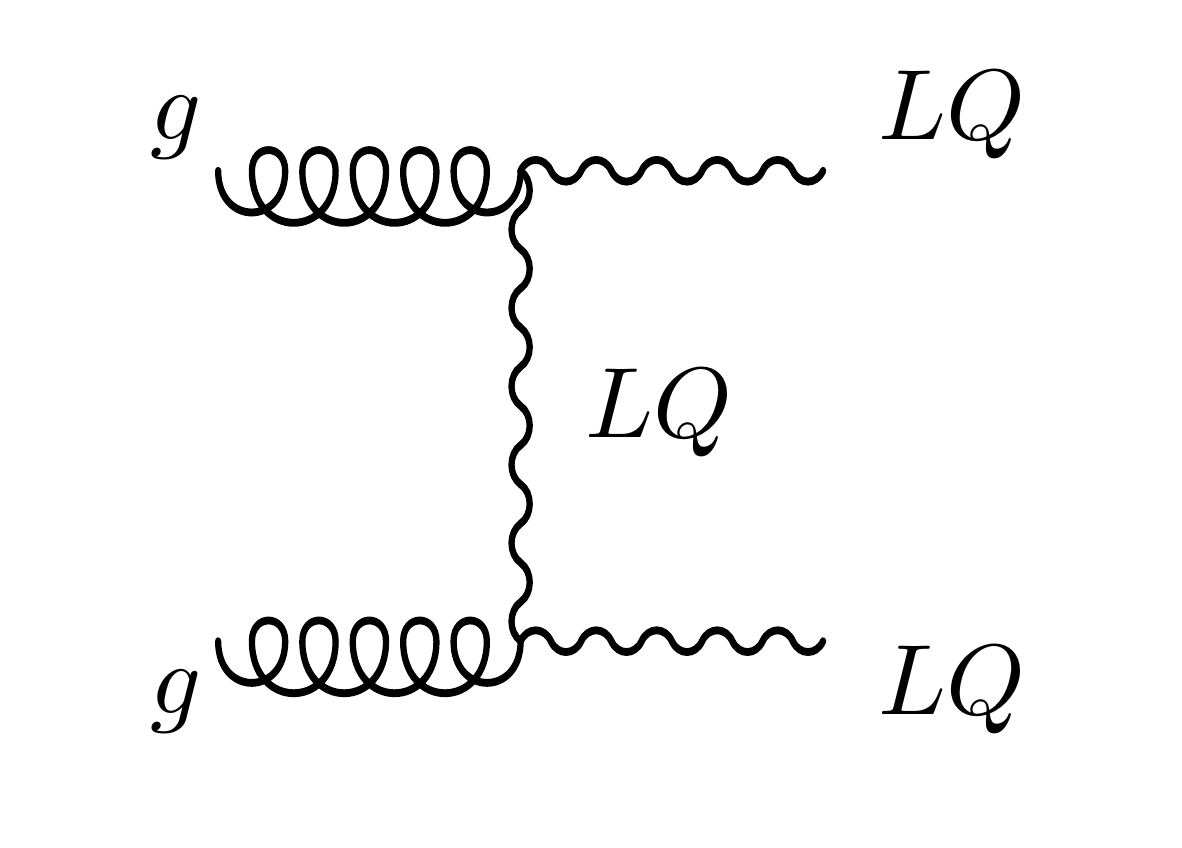}
        \caption{}
    \end{subfigure}
    \hfill
    %non-resonant Leptoquark mediation Diagram
    \begin{subfigure}[b]{0.23\textwidth}
        \includegraphics[width =  1.1\textwidth]{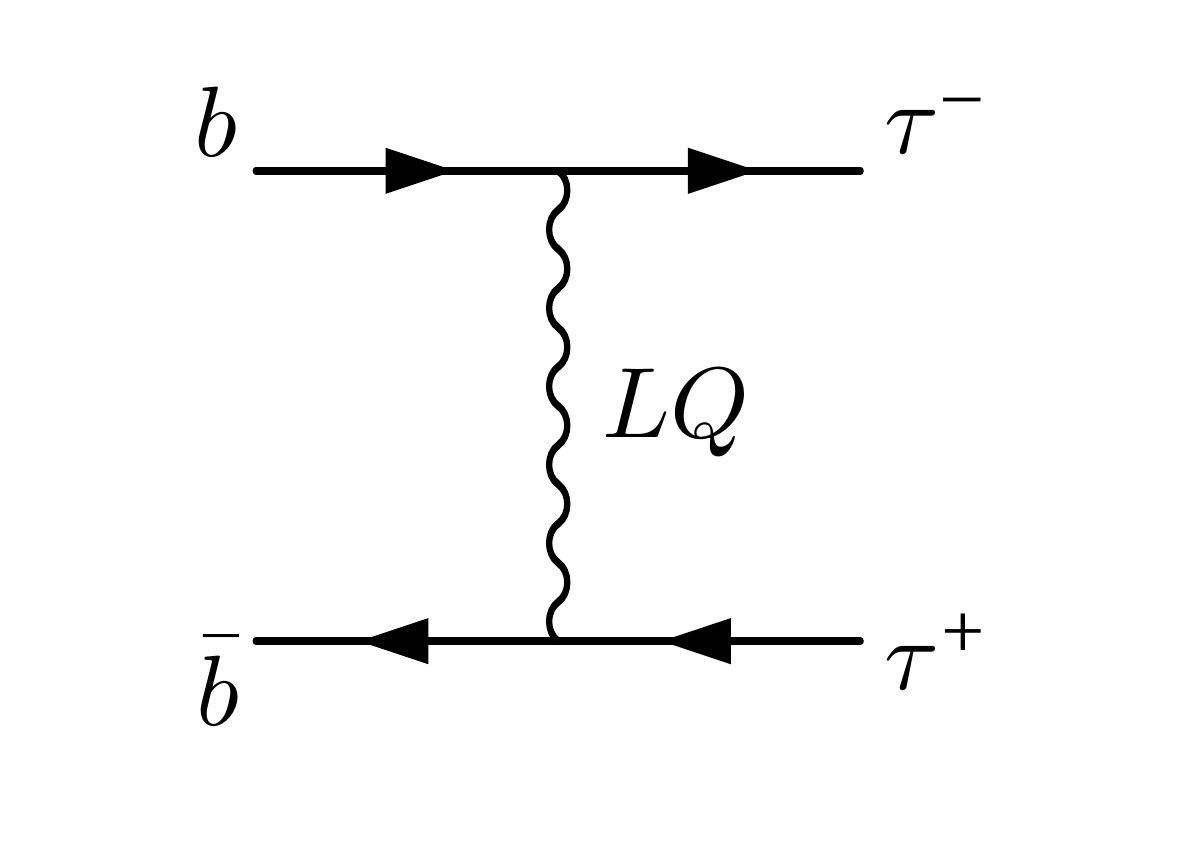}
        \caption{}
    \end{subfigure}
    \caption{Representative Feynman diagrams of single (a), pair  (b), and non-resonant (c) production leptoquarks in proton-proton collision experiments. In single and pair production, the diagrams shown involve t-channel LQ exchange, dominant for lower LQ mass. However, for larger mass there exist s-channel diagrams featuring a virtual bottom quark and gluon, respectively.}
    \label{fig:feynmp-prod-channels}
\end{figure}

Of the searches above, we find~\cite{CMS:2020wzx} particularly interesting. Here, the CMS collaboration explores signals corresponding to $\tq\,\nu\,\bq\,\tau$ and $\tq\,\nu\,\tau$ final states, with $137 \fb^{-1}$ of proton-proton ($\mathrm{p}\,\mathrm{p}$) collision data. The former is motivated by $\lq$ pair production, with one $\lq$ decaying into $\tq\,\nu$ and the other into $\bq\,\tau$, while the latter arises from a single $\lq$ produced in association with a $\tau$, with a subsequent $\lq$ decay into $\tq\,\nu$ (see Figure~\ref{fig:feynmp-prod-channels} for the corresponding diagrams). From the combination of both production channels, the search excludes $U_1$ masses under $1.3-1.7 \tev$, with this range depending on the $U_1$ coupling to gluons and on its coupling $g_U$ in the $\bq_L\,\tau_L$ vertex.

What makes this search particularly attractive is that, for the first time, an LHC collaboration directly places (mass dependent) bounds on $g_U$. This is important, since having information on this parameter is crucial in order to understand if the $U_1$ is really responsible for the $R_{D^{(*)}}$ anomaly. The inclusion of the single-$\lq$ production mode is important, since its cross-section is directly proportional to $g^2_U$. However, as can be seen in Figure~6 of~\cite{CMS:2020wzx}, the current constraints are dominated by pair production, with single-$\lq$ production playing a subleading role. While this is expected~\cite{Schmaltz:2018nls}, it still leads us to ponder the possibility of improving the sensitivity of LHC searches to single-$\lq$ production, and thus on achieving better constraints on $g_U$. Other complementary and similar searches to~\cite{CMS:2020wzx} were carried out by both ATLAS~\cite{ATLAS_7A} and CMS~\cite{LQS_CMS_2022_results_comparison}.

It is also well known, though, that searches for an excess in the high-$\pt$ tails of $\tau$ lepton distributions can strongly probe $g_U$, up to very large $\lq$ masses. Indeed, as shown in~\cite{Faroughy:2016osc,GINO_2022}, the new physics effective operators contributing to $R_D{^{(*)}}$ also contribute to an enhancement in the $\mathrm{p}\,\mathrm{p}\to\tau\tau$ production rates. This has motivated a large number of recasts~\cite{Angelescu:2018tyl,Schmaltz:2018nls,Baker:2019sli,Bhaskar:2021pml,Angelescu:2021lln,Cornella:2021sby,Allwicher:2022gkm,Haisch:2022afh,GINO_2022}, as well as a CMS search explicitly providing constraints in terms of $U_1$~\cite{CMS:2022goy}. Nevertheless, it is important to note that for these $\mathrm{p}\,\mathrm{p}\to\tau\tau$ processes, the $\lq$ participates non-resonantly, so contributions to the $\mathrm{p}\,\mathrm{p}\to\tau\tau$ rates and kinematic distributions from non-LQ BSM diagrams containing possible virtual particles, such as a heavy neutral vector boson $\zb'$, could spoil a straightforward interpretation of any possible excess~\cite{Baker:2019sli}. Thus, it is also necessary to understand how the presence of other virtual particles can affect the sensitivity of an analysis probing $g_U$.

In this work we study the projected $\lq$ sensitivity at the LHC, considering already available $\mathrm{p}\,\mathrm{p}$ data as well as the expected amount of data to be acquired during the High-Luminosity LHC (HL-LHC) runs. We explore a proposed analysis strategy which utilizes a combination of single-, double-, and non-resonant-LQ production, targeting  final states with varying $\tau$-lepton and b-jet multiplicities. 
The studies are performed considering various benchmark scenarios for different $\lq$ masses and couplings, also taking into account distinct chiralities for the third-generation fermions in the $\lq$ vertex. We also assess the impact of a companion $\zb'$, which is typical of gauge models, in non-resonant $\lq$ probes, and find that interference effects can have a significant effect on the discovery reach. We consider this effect to be of high interest, given that non-resonant $\lq$ production can have the largest cross-section, and thus could be an important channel in terms of discovery potential.

An important aspect of this work is that the analysis strategy is developed using a machine learning (ML) algorithm based on Boosted Decision Trees (BDT)\cite{friedman_greedy_2001}. The output of the event classifier is used  to perform a profile-binned likelihood test to extract the overall signal significance for each model considered in the analysis. The advantage of using BDTs and other ML algorithms has been demonstrated in several experimental and phenomenological studies~\cite{Ai:2022qvs,Biswas:2018snp,ATLAS:2017fak,Chigusa:2022svv,Chung:2020ysf,Feng:2021eke,ttZprime}. In our studies, we find that the BDT algorithm gives sizeable improvement in signal significance.

This paper is organized as follows. In Section~\ref{sec:model} we present our simplified model and review the model parameters which are relevant for solving the $\Bm$-meson anomalies. Section~\ref{sec:strategyandsimulation} describes the details associated with the analysis strategy and the simulation of signal and background samples. Section~\ref{sec:results} contains the results of the study, including the projected sensitivity for different benchmark scenarios considered. Finally, in Section~\ref{sec:discusion} we discuss the implication of our results and prospects for future studies. 

\section{A Simplified Model for the $U_1$ Leptoquark}
\label{sec:model}

Extending the SM with a massive $U_1$ vector $\lq$ is not straightforward, as one has to ensure the renormalizability of the model. Most of the theoretical community has focused on extensions of the Pati-Salam (PS) models which avoid proton decay, such as the scenario found in~\cite{Assad:2017iib}. Other examples include PS models with vector-like fermions~\cite{Calibbi:2017qbu,Blanke:2018sro,Iguro:2021kdw}, the so-called 4321 models~\cite{DiLuzio:2017vat,Greljo:2018tuh,DiLuzio:2018zxy}, the twin PS$^2$ model~\cite{King:2021jeo,FernandezNavarro:2022gst}, the three-site PS$^3$ model~\cite{Bordone:2017bld,Bordone:2018nbg,Fuentes-Martin:2022xnb}, as well as composite PS models~\cite{Gripaios:2009dq,Barbieri:2016las,Barbieri:2017tuq}.

In what follows, we shall restrict ourselves to a simplified non-renormalizable lagrangian, understood to be embedded into a more complete model. The SM is thus extended by adding the following terms featuring the $U_1$ $\lq$:
\begin{eqnarray}
\label{eq:BasicLagrangian}
  \mathcal{L}_{U_1}&=&-\frac{1}{2}U^\dagger_{\mu\nu}U^{\mu\nu}+M_U^2\, U_{1\mu}^\dagger U_1^\mu \nonumber \\
 &&  -ig_s\,U_{1\mu}^\dagger\, T^a\, U_{1\nu}\, G^{a\mu\nu}\!\!-i\frac{2}{3}g'\,U^\dagger_{1\mu}U_{1\nu}B^{\mu\nu} \nonumber \\
 && +\frac{g_U}{\sqrt 2}[U_{1\mu}(\bar Q_3\,\gamma^\mu L_3+\beta_L^{s\tau}\,\bar Q_2\,\gamma^\mu L_3 \nonumber \\  && +\beta_{R}\,\bar b_{R}\,\gamma^\mu \tau_{R}) +{\rm h.c.}] 
\end{eqnarray}
where $U_{\mu\nu}\equiv\mathcal{D}_\mu U_{1\nu}-\mathcal{D}_\nu U_{1\mu}$, and $\mathcal{D}_\mu\equiv\partial_\mu+ig_s T^a G_\mu^a+i\tfrac{2}{3}g'B_\mu$. As evidenced by the second line above, we assume that the $\lq$ has a gauge origin \footnote{The couplings in the second line of Eq.~(\ref{eq:BasicLagrangian}) can be found in the literature as $g_s\to g_s(1-\kappa_U)$ and $g'\to g'(1-\tilde\kappa_U)$, in order to take into account the possibility of an underlying strong interaction.}.

The third and fourth lines in in Eq.~(\ref{eq:BasicLagrangian}) shows the $\lq$ interactions with SM fermions, with coupling $g_U$, which we have chosen as preferring the third generation~\footnote{Before the demise of the $R_{K^{(*)}}$ anomaly~\cite{LHCb:2022qnv,LHCb:2022zom,Greljo:2022jac,Ciuchini:2022wbq}, a $3\times3$ $\beta_L$ matrix would be used instead, with values fitted to solve all $\Bm$ meson anomalies.}. These are particularly relevant for the $\lq$ decay probabilities, as well as for the single-$\lq$ production cross-section. The $\beta_L^{s\tau}$ parameter, which is the $\lq \to s\tau$ coupling in the $\beta_L$ matrix (see footnote), is chosen to be equal to $0.2$, following the fit done in~\cite{Cornella:2021sby}, in order to simultaneously solve the $R_{D^{(*)}}$ anomaly and satisfy the $\mathrm{p}\,\mathrm{p}\to\tau^+\tau^-$ constraints. Although $\beta_L^{s\tau}$ technically alters the single-$\lq$ production cross-section and $\lq$ branching fractions, we have confirmed that a value of $\beta_L^{s\tau} = 0.2$ results in negligible impact on our collider results, and thus is ignored in our subsequent studies.

\begin{figure}[]
\centering
    \begin{subfigure}[b]{.92\linewidth}\hspace{5pt}
    \includegraphics[width=.8\linewidth]{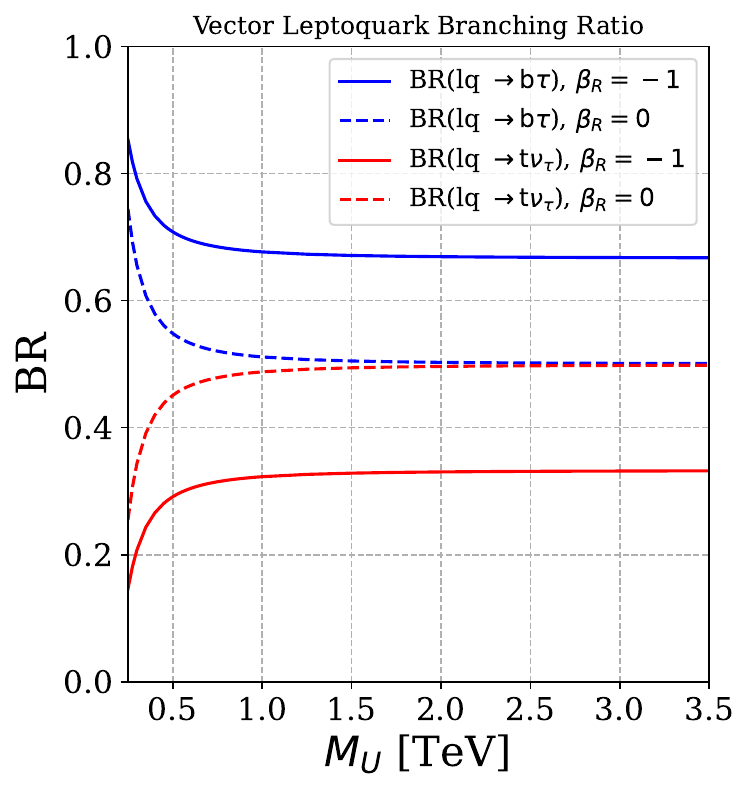}
    \end{subfigure}
    \begin{subfigure}[b]{.94\linewidth}
    \includegraphics[width=.8\linewidth]{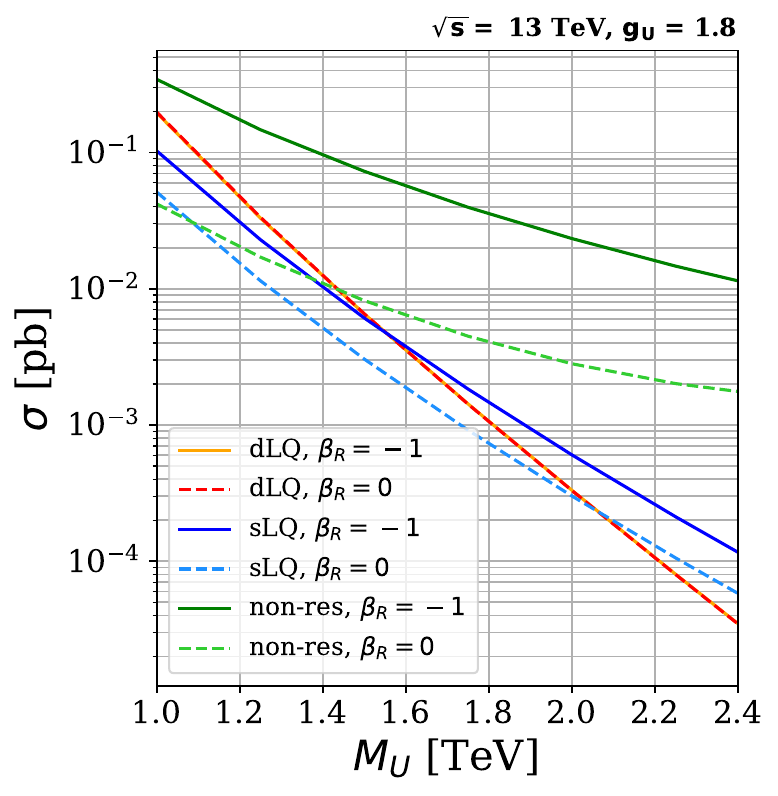}
    \end{subfigure}
    \caption{Top: The $\lq\to\textrm{b}\tau$ and $\lq\to\textrm{t}\nu$ branching ratios for $\beta_{R} = 0$ (solid lines) and $\beta_{R} = -1$ (dashed lines). Bottom: Signal cross-section as a function of the $\lq$ mass, for $\sqrt{ s}=13 \tev$, with $g_U=1.8$. We show single, pair, and non-resonant production, for $\beta_R=-1,\,0$ in solid and dashed lines, respectively.}
\label{fig:branching_ratios}
\end{figure}

The $\lq$ right-handed coupling is modulated with respect to the left-handed one by the $\beta_R$ parameter. The choice of $\beta_R$ is important phenomenologically, as it affects the $\lq$ branching ratios \footnote{Having $\beta_L^{s\tau}$ different from zero also opens new decay channels. These, however, are either suppressed by $\beta_L^{s\tau}$ and powers of $\lambda_{\rm CKM}$. In any case, this effect would decrease ${\rm BR}(\lq \to \bq\,\tau)$ and ${\rm BR}(\lq \to \tq\,\nu)$ by less than $3\%$.}, as well as the single-$\lq$ production cross-section. To illustrate the former, Figure~\ref{fig:branching_ratios} (top) shows the $\lq\to\textrm{b}\tau$ and $\lq\to\textrm{t}\nu$ branching ratios as functions of the $\lq$ mass, for two values of $\beta_R$. For large $\lq$ masses, we confirm that with $\beta_R = 0$ then ${\rm BR}(\lq \to \bq\,\tau) \approx {\rm BR}(\lq \to \tq\,\nu)\approx \tfrac{1}{2}$. However, for $\beta_R = -1$, as was chosen in~\cite{Cornella:2019hct}, the additional coupling adds a new term to the total amplitude, leading to ${\rm BR}(\lq\to \bq\,\tau) \approx \tfrac{2}{3}$. The increase in this branching ratio can thus weaken bounds from $\lq$ searches targeting decays into $\tq\,\nu$ final states, which motivates exploring the sensitivity in b$\tau$ final states exclusively. Note that although a ${\rm BR}(\lq\to \bq\,\tau) \approx 1$ scenario is possible by having the $\lq$ couple exclusively to right-handed currents (i.e, $g_U\to0$, but $g_U\beta_R\not=0$), it does not solve the observed anomalies in the $R_{D^{(*)}}$ ratios. Therefore, although some LHC searches assume ${\rm BR}(\lq\to \bq\,\tau) = 1$, we stress that in our studies we assume values of the model parameters and branching ratios that solve the $R_{D^{(*)}}$ ratios.

To further understand the role of $\beta_R$ at colliders, Figure~\ref{fig:branching_ratios} (bottom) shows the cross-section for single-$\lq$ (s$\lq$), double-$\lq$ (d$\lq$), and non-resonant (non-res) production, as a function of mass and for a fixed coupling $g_{U} = 1.8$, assuming $\mathrm{p}\,\mathrm{p}$ collisions at $\sqrt{s} = 13$ $\tev$. We note that this benchmark scenario with $g_{U}=1.8$ results in a $\lq\to\textrm{b}\tau$ decay width that is $<$5\% of the $\lq$ mass, for mass values from 250 $\gev$ to 2.5 $\tev$. In the Figure, we observe that, since d$\lq$ production is mainly mediated by events from quantum chromodynamic processes, the choice of $\beta_R$ does not affect the cross-section. However, for  s$\lq$ production, a non-zero value for $\beta_R$ increases the cross-section by about a factor of 2 and by almost one order of magnitude in the case of non-res production. These results shown in Figure~\ref{fig:branching_ratios} are easily understood by considering the diagrams shown in Figure~\ref{fig:feynmp-prod-channels}. The $\lq$ mass value where the s$\lq$ production cross-section exceeds the d$\lq$ cross-section depends on the choice of $g_U$. 
 
We also note that to solve the $R_{D^{(*)}}$ anomaly, the authors of~\cite{Cornella:2021sby} point out that the wilson coefficient $C_U\equiv g^2_U\,v^2_{SM}/(4\,M^2_U)$ is constrained to a specific range of values, and this range depends on the value of the $\beta_{R}$ parameter. Therefore, the allowed values of the coupling $g_{U}$ depend on $M_{U}$ and $\beta_{R}$, and thus our studies are performed in this multi-dimensional phase space.

As noted in section~\ref{sec:intro}, we study the role of a $\zb'$ boson in $\mathrm{p}\,\mathrm{p}\to\tau\tau$ production. The presence of a $\zb'$ boson in $\lq$ models has been justified in various papers, for example, in~\cite{Baker:2019sli}. The argument is that minimal extensions of the SM which include a massive gauge $U_1$ LQ, uses the gauge group $SU(4)\times SU(3)^{\prime}\times SU(2)_L \times U(1)_{T_R^3}$. Such an extension implies the presence of an additional massive boson, $\zb^{\prime}$, and a color-octet vector, $G'$, arising from the spontaneous symmetry breaking into the SM \footnote{Naively, the LQs are associated to the breaking of $SU(4)\to SU(3)_{[4]}\times U(1)_{B-L}$, the $G'$ arises from $SU(3)_{[4]}\times SU(3)'\to SU(3)_c$, and the $Z'$ comes from the breaking of $U(1)_{B-L}\times U(1)_{T_R^3}\to U(1)_Y$. Notice that the specific pattern of breaking, and the relations between the masses and couplings, are connected to the specific scalar potential used.}.  The $\zb'$ in particular can play an important role in the projected $\lq$ discovery reach, as it can participate in $\mathrm{p}\,\mathrm{p}\to\tau\tau$ production by s-channel exchange, both resonantly and as a virtual mediator. To study the effect of a $\zb'$ on the $\mathrm{p}\,\mathrm{p}\to\tau\tau$ production cross-sections and kinematics, we extend our benchmark Lagrangian in Eq.~(\ref{eq:BasicLagrangian}) with further non-renormalizable terms involving the $\zb'$. Accordingly, we assume the $\zb'$ only couples to third-generation fermions. Our simplified model is thus extended by:
\begin{eqnarray}
    \label{eq:BasicLagrangianZp}
        \mathcal{L}_{Z^{\prime}}&= & -\frac{1}{4} Z_{\mu \nu}^{\prime} Z^{\prime \mu \nu}+\frac{1}{2} M_{Z^{\prime}}^2 Z_\mu^{\prime} Z^{\prime \mu} \nonumber \\
        && + \frac{g_{Z^{\prime}}}{2 \sqrt{6}} Z^{\prime \mu} (\zeta_q \bar{Q}_3 \gamma_\mu Q_3 \nonumber +\zeta_t \bar{t}_R \gamma_\mu t_R \\
        &&  +\zeta_b \bar{b}_R \gamma_\mu b_R-3 \zeta_{\ell} \bar{L}_3 \gamma_\mu L_3-3 \zeta_\tau \bar{\tau}_R \gamma_\mu \tau_R)
\end{eqnarray}
where the constants $M_{\zb^{\prime}}$, $g_{Z^{\prime}}$, $\zeta_q $, $\zeta_t $, $\zeta_b$, $\zeta_{\ell}$, $\zeta_\tau$, are model dependent.

\begin{figure}[]
\centering
    \begin{subfigure}[b]{.94\linewidth}
    \includegraphics[width=.8\linewidth]{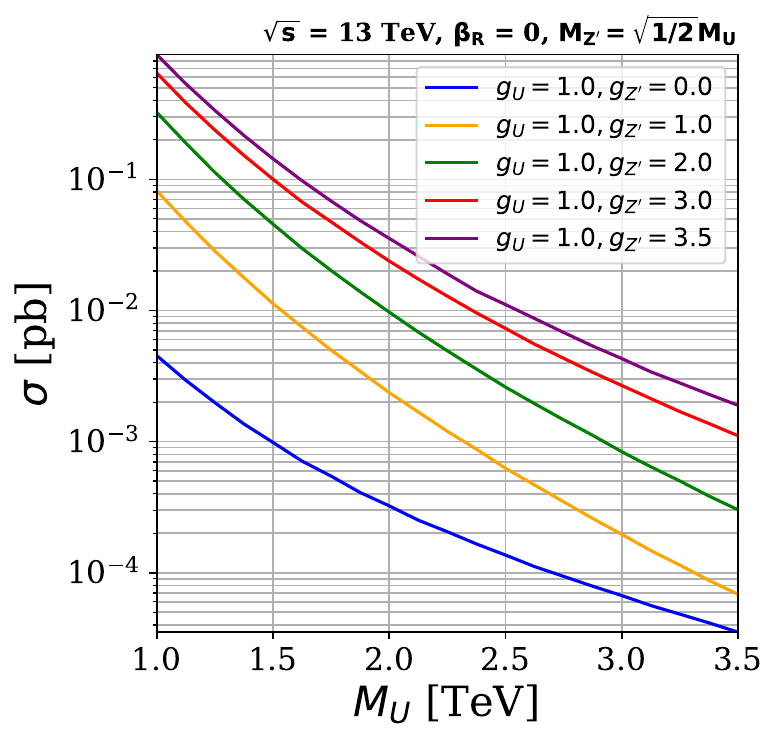}
    \end{subfigure}
    \begin{subfigure}[b]{.94\linewidth}
    \includegraphics[width=.8\linewidth]{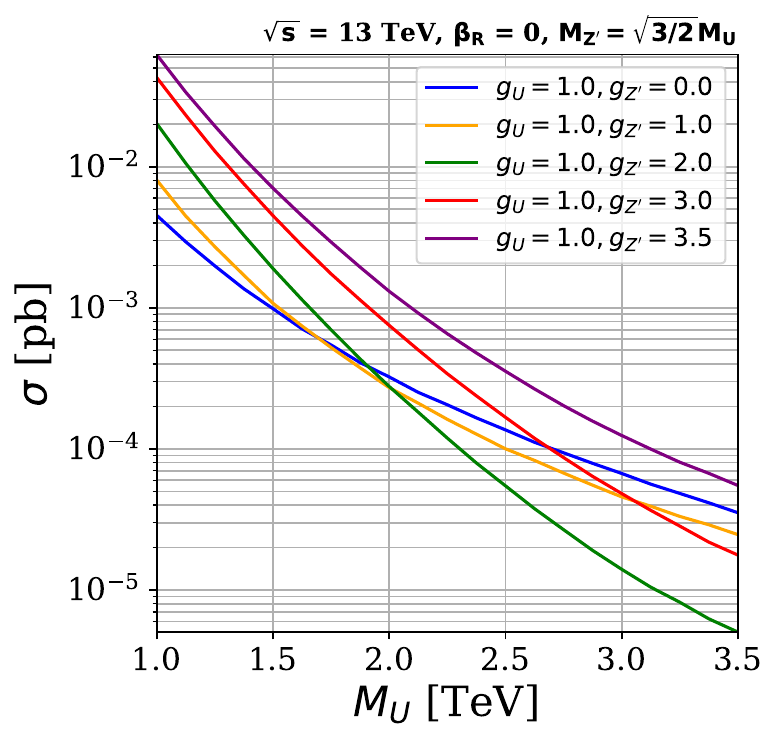}
    \end{subfigure}
    \caption{$\tau \tau$ cross-section as a function of the $\lq$ mass for different values of $g_U$ and $g_{\zb^{\prime}}$. The estimates are performed at $\sqrt s=13 \tev$, $\beta_R=0$,  $M_{\zb^{\prime}} = \sqrt{1/2} M_{U}$ (top), and $M_{\zb^{\prime}} = \sqrt{3/2} M_{U}$ (bottom).}
\label{fig:xsinterference}
\end{figure}

We study two extreme cases for the $\zb'$ mass, following~\cite{GINO_PhysRevD.102.115015}, namely $M_{\zb'} = \sqrt{\tfrac{1}{2}}M_U<M_U$ and $M_{\zb'} = \sqrt{\tfrac{3}{2}}M_U>M_U$. We also assume the $\lq$ and $\zb'$ are uniquely coupled to left-handed currents, i.e. $\zeta_q=\zeta_\ell= 1$ and $\zeta_t=\zeta_b=\zeta_\tau=0$. With these definitions, Figure~\ref{fig:xsinterference} shows the effect of the $\zb'$ on the $\tau\tau$ production cross-section, considering $g_U = 1$, $\beta_R=0$, and different $g_{\zb^{\prime}}$ couplings. On the top panel, the cross-sections corresponding to the cases where $M_{\zb'} = \sqrt{\tfrac{1}{2}}M_U$ are shown. As expected, the $\tau\tau$ production cross-section for the inclusive case (i.e., $g_{\zb'} \neq 0$) is larger than that for the $\lq$-only non-res process ($g_{\zb'} = 0$, depicted in blue). This effect increases with $g_{\zb'}$ and, within the evaluated values, can exceed the $\lq$-only cross-section by up to two orders of magnitude. In contrast, a more intricate behaviour can be seen in the bottom panel of Figure~\ref{fig:xsinterference}, which corresponds to $M_{\zb'} = \sqrt{\tfrac{3}{2}}M_U$. Here, for low values of $M_U$, a similar increase in the cross-section is observed. However, for higher values of $M_U$, the inclusive $\mathrm{p}\,\mathrm{p}\to\tau\tau$ cross-section is smaller than the $\lq$-only $\tau\tau$ cross-section. This behaviour suggests the presence of a dominant destructive interference at high masses, leaving its imprint on the results.

\begin{figure}[]
\centering
    \begin{subfigure}[b]{.94\linewidth}
    \includegraphics[height=6cm, width=6.0cm]{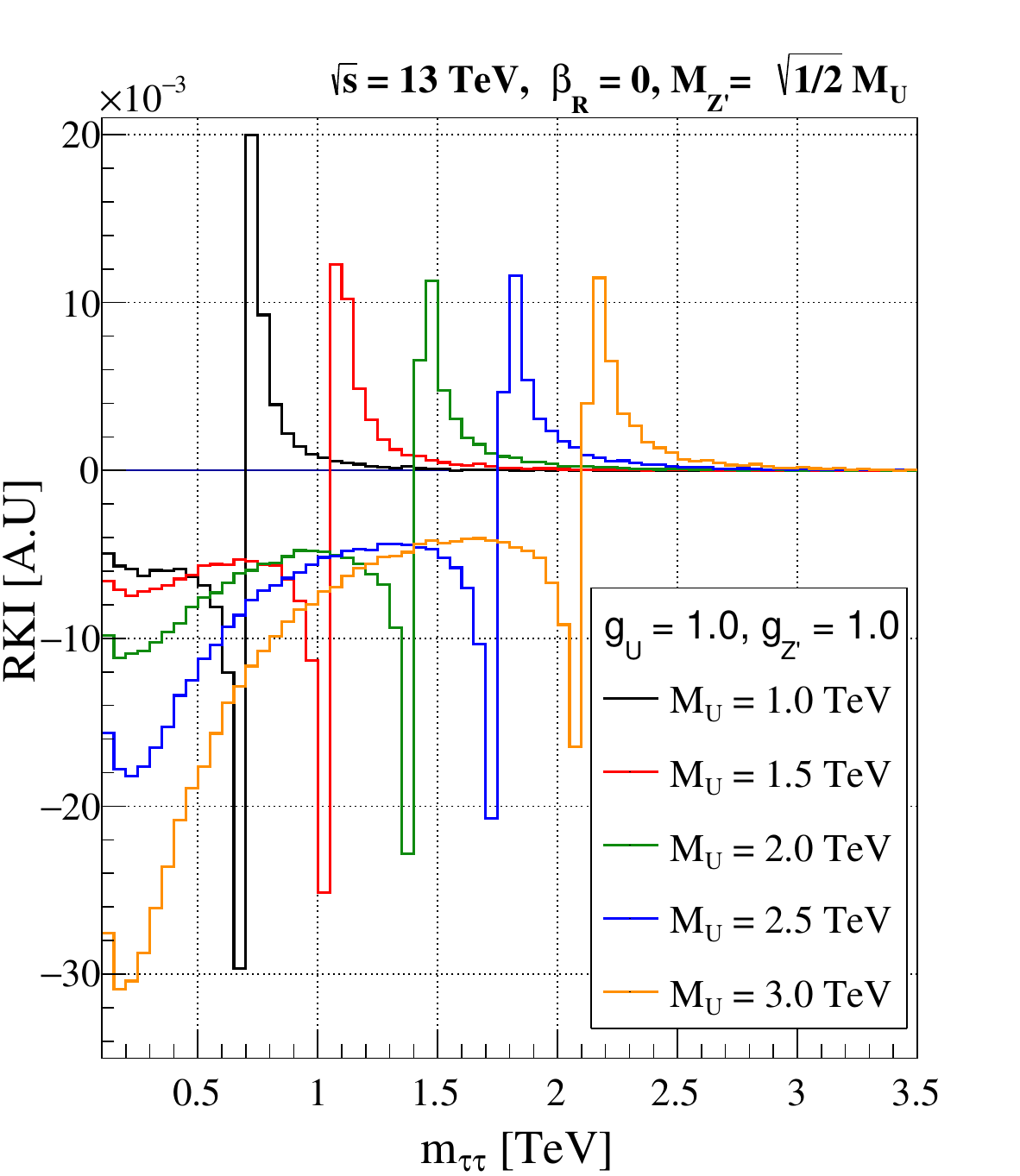}
    \end{subfigure}
    \begin{subfigure}[b]{.94\linewidth}
    \includegraphics[height=6cm, width=6cm]{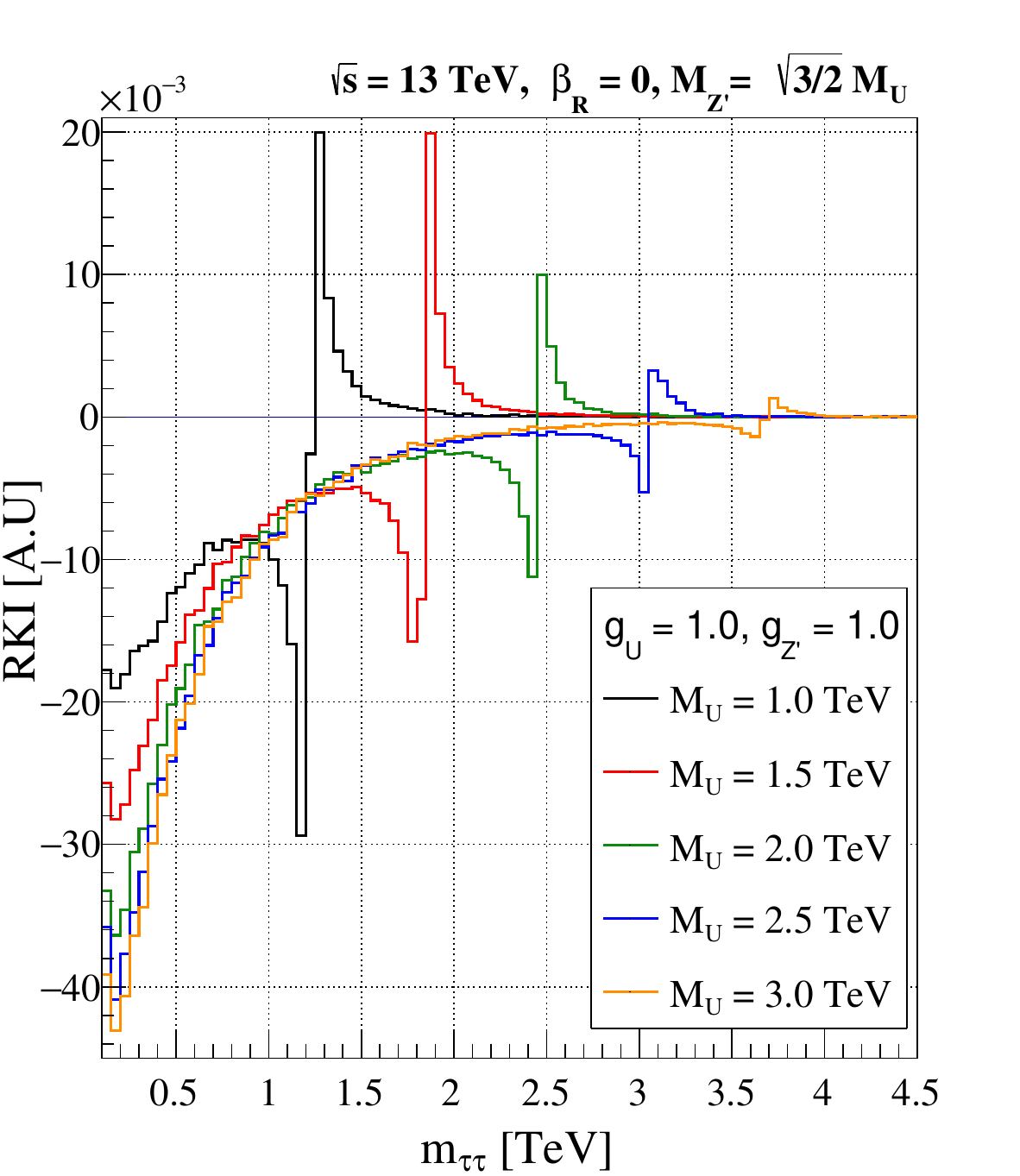}
    \end{subfigure}
    \caption{The relative kinematic interference (RKI), as a function of the reconstructed mass of two taus, for different $\lq$ masses. The studies are performed assuming $\sqrt s=13 \tev$, $\beta_R=0$, $g_U = 1.0$, $g_{\zb^{\prime}} =1.0$, $M_{\zb^{\prime}} = \sqrt{1/2} M_{U}$ (top), and $M_{\zb^{\prime}} = \sqrt{3/2} M_{U}$ (bottom).
    }    
\label{fig:interference}
\end{figure}
In order to further illustrate the effect, Figure~\ref{fig:interference} shows the relative kinematic interference ($\mathrm{RKI}$) as a function of the reconstructed invariant mass $m_{\tau\tau}$, for $g_{\zb^{\prime}} = 1$ and varying values of $M_U$. The RKI parameter is defined as
\begin{equation}
    \mathrm{RKI}(m_{\tau\tau})=\frac{1}{\sigma_{\lq+\zb'}}\left[\frac{d\sigma_{\lq+\zb'}}{dm_{\tau\tau}}-\left(\frac{d\sigma_{\lq}}{dm_{\tau\tau}}+\frac{d\sigma_{\zb'}}{dm_{\tau\tau}}\right)\right],
\end{equation}
where $\sigma_{X}$ is the production cross-section arising due to contributions from $X$ particles. For example, $\sigma_{\lq+\zb'}$ represents the inclusive cross-section where both virtual $\lq$ and s-channel $\zb'$ exchange contribute. For both cases, we can observe the presence of deep valleys in the RKI curves when $m_{\tau\tau}\to0$, indicating destructive interference between the $\lq$ and the $\zb'$ contributions. This interference generates a suppression of the differential cross-section for lower values of $m_{\tau\tau}$ and, therefore, in the integrated cross-section. 
 
The observed interference effects are consistent with detailed studies on resonant and non-res $\mathrm{p}\,\mathrm{p}\to\tq \bar{\tq}$ production, performed in reference~\cite{Djouadi:2019cbm}.

\section{$\lq$ search strategy and simulation}
\label{sec:strategyandsimulation}

Our proposed analysis strategy utilizes single-$\lq$ (i.e.\ $\mathrm{p}\,\mathrm{p}\to \tau\,\lq$), double-$\lq$ (i.e.\ $\mathrm{p}\,\mathrm{p}\to \lq\,\lq$), and non-resonant $\lq$ production (i.e.\ $\mathrm{p}\,\mathrm{p}\to \tau\tau$) as shown in Figure~\ref{fig:feynmp-prod-channels}. At leading order in $\alpha_s$, since we focus on  $U_1\to \bq\,\tau$ decays, the s$\lq$ process results in the $\textrm{b}\tau\tau$ mode, the d$\lq$ process results in the $\textrm{bb}\tau\tau$ mode, and the non-res process results in the $\tau\tau$ mode. Therefore, in all cases we obtain two $\tau$ leptons, with either 0, 1, or 2 b jets. The $\tau$ leptons decay to hadrons ($\tau_{\textrm{h}}$) or semi-leptonically to electrons or muons ($\tau_{\ell}$, $\ell = \textrm{e}$ or $\mu$). To this end, we study six final states: $\tau_{\mathrm h} \tau_{\mathrm{h}/\ell}$, $\bq\,\tau_{\mathrm h} \tau_{\mathrm{h}/\ell}$, and $\bq \bq\,\tau_{\mathrm h} \tau_{\mathrm{h}/\ell}$, which can be naively associated to non-res, s$\lq$ and d$\lq$ production, respectively.
Nevertheless, experimentally it is possible for $\bq$ jets to not be properly identified or reconstructed, leading, for instance, to a fraction of d$\lq$ signal events falling into the $\bq\,\tau_{\mathrm h} \tau_{\mathrm{h}/\ell}$ and $\tau_{\mathrm h}\tau_{\mathrm{h}/\ell}$ categories. Similarly, soft jets can fake $\bq$ jets, such that non-res processes can contribute to the $\bq\,\tau_{\mathrm h} \tau_{\mathrm{h}/\ell}$ and $\bq\bq\,\tau_{\mathrm h} \tau_{\mathrm{h}/\ell}$ final states. This kind of signal loss and mixing is taken into account in our analysis\footnote{Note that further signal mixing can also occur at the event generation level by including terms at larger order in $\alpha_s$. For example, in the non-res diagram in Figure~\ref{fig:feynmp-prod-channels}, one of the initial $\bq$ could come from a $g\to \bq\bar{\bq}$ splitting, leading to non resonant production of $\bq\,\tau_{\mathrm h} \tau_{\mathrm{h}/\ell}$. Simulating and studying the role of such NLO contributions is outside the scope of this work.}.

The contributions of signal and background events are estimated using Monte Carlo (MC) simulations. We implemented the $U_1$ model from~\cite{Baker:2019sli}, adjusted to describe the lagrangian in Equations~\eqref{eq:BasicLagrangian} and ~\eqref{eq:BasicLagrangianZp}, using \texttt{FeynRules} (v2.3.43) ~\cite{Christensen:2008py,Alloul:2013bka}. The branching ratios and cross-sections have been calculated using \texttt{MadGraph5\_aMC} (v3.1.0) \cite{Alwall:2014bza, Alwall:2014hca}, the latter at leading order in $\alpha_s$. The corresponding samples are generated considering $\mathrm{p}\,\mathrm{p}$ collisions at  $\sqrt{s}=13 \tev$ and $\sqrt{s}=13.6 \tev$.  All samples are generated using the NNPDF3.0 NLO~\cite{NNPDF:2014otw} set for parton distribution functions (PDFs) and using the full amplitude square SDE strategy for the phase-space optimization due to strong interference effects with the $\zb'$ boson. Parton level events are then interfaced with the \texttt{PYTHIA} (v8.2.44)~\cite{Sjostrand:2014zea} package to include parton fragmentation and hadronization processes, while \texttt{DELPHES} (v3.4.2)~\cite{deFavereau:2013fsa} is used to simulate detector effects, using the input card for the CMS detector geometric configurations, and for the performance of particle reconstruction and identification.

At parton level, jets and leptons are required to have a minimum transverse momentum ($\pt$) of $20 \gev$, while $\bq$ jets are required to have a minimum $\pt$ of $30 \gev$. Additionally, we constrain the pseudorapidity ($\eta$) to $|\eta| < 2.5$ for $\bq$ jets and leptons, and $|\eta| < 5.0$ for jets. The production cross-sections shown in the bottom panel of Figures~\ref{fig:branching_ratios} and~\ref{fig:xsinterference} are obtained with the aforementioned selection criteria. 

\begin{table}[t]
    \begin{tabular}{|c|cccccc|}
\hline
\multirow{2}{*}{\textbf{Variable}} & \multicolumn{6}{c|}{\textbf{Threshold}} \bigstrut\\ 
\cline{2-7} 
                                   & \multicolumn{1}{c|}{$\tau_{h} \tau_{h}$} & \multicolumn{1}{c|}{$\bq \tau_{h} \tau_{h}$} & \multicolumn{1}{c|}{$\bq \bq \tau_{h} \tau_{h}$} & \multicolumn{1}{c|}{$\tau_{h} \tau_{\ell}$} & \multicolumn{1}{c|}{$\bq \tau_{h} \tau_{\ell}$} & $\bq \bq \tau_{h} \tau_{\ell}$ \bigstrut\\ \hline \hline

$N(\bq)$                             & \multicolumn{1}{c|}{= 0} & \multicolumn{1}{c|}{= 1} & \multicolumn{1}{c|}{$\geq 2$} & \multicolumn{1}{c|}{= 0}& \multicolumn{1}{c|}{= 1} & $\geq 2$ \bigstrut\\ \hline
$\pt(\bq)$                         & \multicolumn{1}{c|}{-} & \multicolumn{2}{c|}{$\geq 30 \gev$ } & \multicolumn{1}{c|}{-} & \multicolumn{2}{c|}{$\geq 30 \gev$ } \bigstrut\\ \hline
$|\eta(\bq)|$                        & \multicolumn{1}{c|}{-} &\multicolumn{2}{c|}{$\leq 2.4$} & \multicolumn{1}{c|}{-} &\multicolumn{2}{c|}{$\leq 2.4$} \bigstrut\\ \hline
                                  
$N (\ell)$                         & \multicolumn{3}{c|}{= 0}  & \multicolumn{3}{c|}{= 1} \bigstrut\\ \hline
$\pt(\el)$                         & \multicolumn{3}{c|}{-} & \multicolumn{3}{c|}{$\geq 35 \gev$} \bigstrut\\ \hline
$\pt(\mu)$                       & \multicolumn{3}{c|}{-} & \multicolumn{3}{c|}{$\geq 30 \gev$} \bigstrut\\ \hline
$|\eta(\ell)|$                     & \multicolumn{3}{c|}{-} & \multicolumn{3}{c|}{$\leq 2.4$} \bigstrut\\ \hline

$N(\tau_{h})$                      & \multicolumn{3}{c|}{ $= 2$} & \multicolumn{3}{c|}{= 1} \bigstrut\\ \hline
$\pt(\tau_h)$                    & \multicolumn{6}{c|}{$\geq 50$ GeV} \bigstrut\\ \hline
$|\eta(\tau_h)|$                   & \multicolumn{6}{c|}{$\leq 2.3$} \bigstrut\\ \hline
$\Delta R(p_{i}, p_{j})$           & \multicolumn{6}{c|}{$\geq 0.3$} \bigstrut\\ \hline

\end{tabular}

    \caption{Preliminary event selection criteria used to filter events before feeding them to the BDT algorithm. A $\Delta R(p_{i}, p_{j}) > 0.3$ requirement is imposed between all pairs of reconstructed particle candidates $p_{i}, p_{j}$. 
    }
    \label{table:selection_channels}
\end{table}
Table~\ref{table:selection_channels} shows the preliminary event selection criteria for each channel at analysis level. The channels are divided based on the multiplicity of $\bq$ jets, $N(\bq)$, number of light leptons, $N(\ell)$, number of hadronic tau leptons, $N(\tau_{\mathrm h})$, and kinematic criteria based on $\eta$, $\pt$ and spatial separation of particles in the detector volume $(\Delta R = \sqrt{(\Delta \eta)^{2} + (\Delta \phi)^{2}})$. The minimum $\pt$ thresholds for leptons are chosen  following references~\cite{CMS:2020wzx, CMS:2022goy, ATLAS:2021oiz}, based on experimental constrains associated to trigger performance. Following reference~\cite{CMS_BTV2016}, we use a flat identification efficiency for $\bq$ jets of 70\% across the entire $\pt$ spectrum with misidentification rate of 1\%. These values correspond with the  ``medium working point'' of the CMS algorithm to identify $\bq$ jets, known as DeepCSV. We also explored the ``Loose'' (``Tight'') working point using an efficiency of 85\% (45\%) and mis-identification rate of 10\% (0.1\%). The  ``medium working point'' was selected as it gives the best signal significance for the analysis. 

For the performance of $\tau_{\textrm{h}}$ identification in DELPHES, we consider the latest technique described in~\cite{CMS_DeepTau}, which is based on a deep neural network (i.e. DeepTau) that combines variables related to isolation and $\tau$-lepton lifetime as input to identify different $\tau_{\textrm{h}}$ decay modes. Following~\cite{CMS_DeepTau}, we consider three possible DeepTau ``working points'': (i) the ``Medium'' working point of the algorithm, which gives a 70\% $\tau_{\textrm{h}}$-tagging efficiency and 0.5\% light-quark and gluon jet mis-identification rate; (ii) the ``Tight'' working point, which gives a 60\% $\tau_{\textrm{h}}$-tagging efficiency and 0.2\% light-quark and gluon jet mis-identification rate; and (iii) the ``VTight'' working point, which gives a 50\% $\tau_{\textrm{h}}$-tagging efficiency and 0.1\% light-quark and gluon jet mis-identification rate. Similar to the choice of $\textrm{b}$-tagging working point, the choice of $\tau_{\textrm{h}}$-tagging working point is determined through an optimization process which maximizes discovery reach. The ``Medium'' working point was ultimately shown to provide the best sensitivity and therefore chosen for this study. For muons (electrons), the assumed identification efficiency is 95\% (85\%), with a 0.3\% (0.6\%) mis-identification rate~\cite{CMS-PAS-FTR-13-014,CMS_MUON_17001,CMS_EGM_17001}.

After applying the preliminary selection criteria, the primary sources of background are production of top quark pairs ($\tq\bar{ \tq}$), and single-top quark processes (single $\tq$), followed by production of vector bosons with associated jets from initial or final state radiation ($V$+jets), and pair production of vector bosons ($VV$). The number of simulated MC events used for each sample is shown in Table~\ref{table:MC_events}. 

\begin{table}[]
    %Vertical

% \begin{table}[!h]
% \begin{tabular}{|c|c|}
% \hline
% \textbf{Sample} & \textbf{$\mathbf{N_{events}}$} \\ \hline
% $\tq \bar{\tq}$                  & 24.307.250                      \\ \hline
% single $\tq$                         & 11.500.000                      \\ \hline
% $VV$                      & 32.350.000                      \\ \hline
% $V+$jets                       & 39.448.395                    \\ \hline
% signals       & 600.000                        \\ \hline
% \end{tabular}
%     \caption{Initial number of events produced for the signal and background samples. \JJP{Again, long and thin. Maybe transpose the table? That is, instead of $6\times2$ make it $2\times6$?}}
%     \label{table:MC_events}
% \end{table}

%Horizontal

\begin{tabular}{|c|c|c|c|c|c|}
\hline
\textbf{Sample}               & $\tq \bar{\tq}$ & single $\tq$ & $VV$       & $V+$jets   & signals \bigstrut\\ \hline \hline
\textbf{$\mathbf{N_{events}}\times 10^{-6}$} & 24.31      & 11.50   & 32.35 & 39.45 & 0.60 \bigstrut\\ \hline
\end{tabular}

    \caption{The number of simulated events for the signal and background samples.}
    \label{table:MC_events}
\end{table}

We use two different sets of signal samples. The first set includes various $\set{M_{U},g_{U}}$ scenarios, for two different values of $\beta_R\in \set{0,-1}$. We generate signal samples for $M_{U}$ values between 250 GeV and 5000 GeV, in steps of 250 GeV. The considered $g_{U}$ coupling values are between 0.25 and 3.5, in steps of 0.25. Although the signal cross-sections depend on both $M_{U}$ and $g_{U}$, the efficiencies of our selections only depend on $M_{U}$ (for all practical purposes) since the decay widths are relatively small compared to the mass of $M_{U}$ ($\frac{\Gamma_{U}}{M_{U}} < 5$\%), and thus more sensitive to experimental resolution. In total there are 280 $\set{M_{U},g_{U},\beta_{R}}$ scenarios simulated for this first set of signal samples, and for each of these scenarios two subsets of samples are generated, which are used separately for the training and testing of the machine learning algorithm. The second set of signal samples is used to evaluate interference effects between $\lq$s and the $\zb^{\prime}$ bosons in non-res production. Using benchmark values $g_U=1.8$ and $\beta_R=0$, we consider various $\set{M_{U},g_{\zb^{\prime}}}$ scenarios for two different $\zb^{\prime}$ mass hypotheses, $\left(M_{\zb'}/M_U\right)^2 \in \Set{\tfrac{1}{2},\tfrac{3}{2}}$. The $M_{U}$ values vary between 500 GeV and 5000 GeV, in steps of 250 GeV. The $g_{\zb^{\prime}}$ coupling values are between 0.25 and 3.5, in steps of 0.25. Therefore, in total there are 280 $\set{M_{U},g_{\zb^{\prime}},\left(M_{\zb'}/M_U\right)^2}$ scenarios simulated for this second set of signal samples, and for each of these scenarios a total of $6.0 \times 10^{5}$ MC events are generated.

As noted previously, the simulated signal and background events are initially filtered using selections which are motivated by experimental constraints, such as the geometric constraints of the CMS detector, the typical kinematic thresholds for reconstruction of particle objects, and the available triggers. The remaining events after the preliminary event selection criteria are used to train and execute a BDT algorithm for each signal point in the $\set{M_{U},g_{U}}$ space, in order to maximize the probability to detect signal amongst background events. The BDT algorithm is implemented using the \texttt{scikit-learn}~\cite{pedregosa_scikit-learn_2011} and \texttt{xgboost} (XGB)~\cite{chen_xgboost_2016} python libraries. We use the the \texttt{XGBClassifier} class from the \texttt{xgboost} library, a 10-fold cross validation using the \texttt{scikit-learn} method (\texttt{GridCV}  \footnote{GridCV is a method that allows to find the best combination of hyperparameter values for the model, as this choice is crucial to achieve an optimal performance.}) for a grid in a hyperparameter space with 75, 125, 250, and 500 estimators, maximum depth in 3, 5, 7, 9, as well as learning rates of 0.01, 0.1, 1, and 10. For the cost function, we utilize the default mean square error (\texttt{MSE}). Additionally, we use the tree method based on the approximate greedy algorithm (histogram-optimized), referred to as \texttt{hist}, with a uniform sample method. These choices allow us to maximize the detection capability of the BDT algorithm by carefully tuning the hyperparameters, selecting an appropriate cost function, and utilizing an optimized tree construction method.

For each of the six analysis channels and $\set{M_{U},g_{U}}$ signal point, the binary XGB classifier was trained (tested) with 20\% (80\%) of the simulated events, for each signal and background MC sample. 
%, under the narrow width approximation. 
Over forty kinematic and topological variables were studied as input for the XGB. These included the momenta of b jets and $\tau_{\textrm{h},\ell}$ candidates; both invariant and transverse masses of pairs of $\tau$ objects and of $\textrm{b}\,\tau$ combinations; angular differences between b jets, between $\tau$ objects, and between the $\tau_{\textrm{h},\ell}$ and b jets; and additional variables derived from the missing momentum in the events. After studying correlations between variables and their impact on the performance of the BDT, we found that only eight variables were necessary and responsible for the majority of the sensitivity of the analysis.  
The variable that provides the best signal to background separation is the scalar sum of the $\pt$ of the final state objects ($\tau_{\mathrm h}$, $\tau_{h/\ell}$, and $\bq$ jets) and the missing transverse momentum, referred to as $S_{\mathrm{T}}^{MET}$: 
\begin{equation}
S_{\mathrm{T}}^{MET}=|\vec{p}_T^{\;miss}|+\sum_{\tau_{\mathrm h},\,\tau_{h/\ell},\,\bq}|\vec{p}_T|
\end{equation}
The $S_{\mathrm{T}}^{MET}$ variable has been successfully used in $\lq$ searches at the LHC, since it probes the mass scale of resonant particles involved in the production processes. Other relevant variables include the magnitude of the vectorial difference in $\pt$ between the two lepton candidates ($|\Delta \vec{p}_T |_{\tau_{\textrm{h}} \tau_{\textrm{h}/\ell}}$), the $\Delta R_{\tau_{\textrm{h}} \tau_{\textrm{h}/\ell}}$ separation between them, the reconstructed dilepton mass $m_{\tau_{\textrm{h}}\tau_{\textrm{h}/\ell}}$, and the product of their electric charges ($Q_{\tau_{\textrm{h}}} \times Q_{\tau_{\textrm{h}/\ell}}$). 
We also use the $|\Delta \vec{p}_T|$ between the $\tau_{\mathrm h}$ candidate and $\vec{p}_T^{\,miss}$, and (if applicable) the $|\Delta \vec{p}_T|$ between the $\tau_{\textrm{h}}$ candidate and the leading $\bq$ jet. For the final states including two $\tau_{\textrm{h}}$ candidates, the one with the highest $\pt$ is used.

\begin{figure}[]
    \centering
    \includegraphics[height=4.9cm]{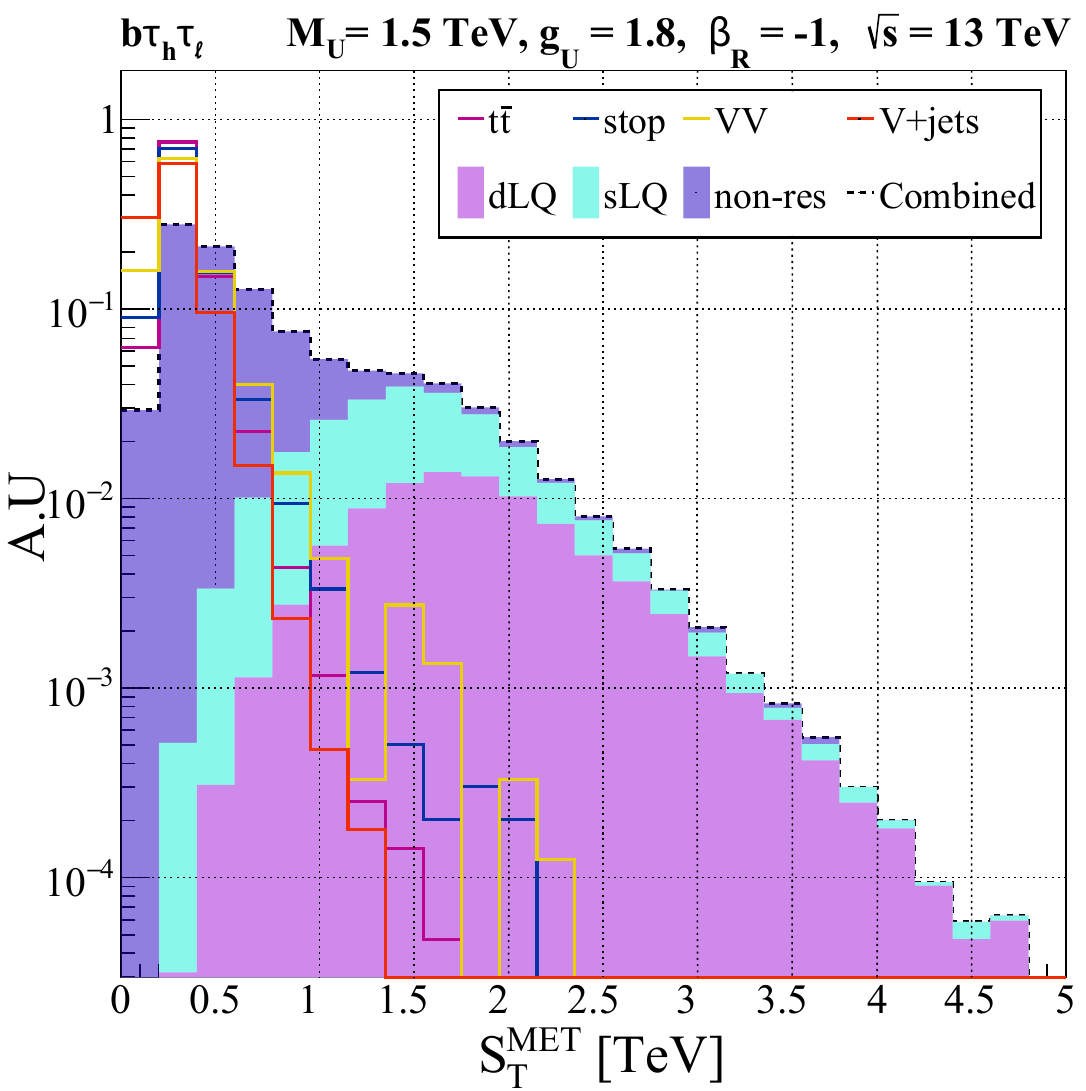}
    \includegraphics[height=4.9cm]{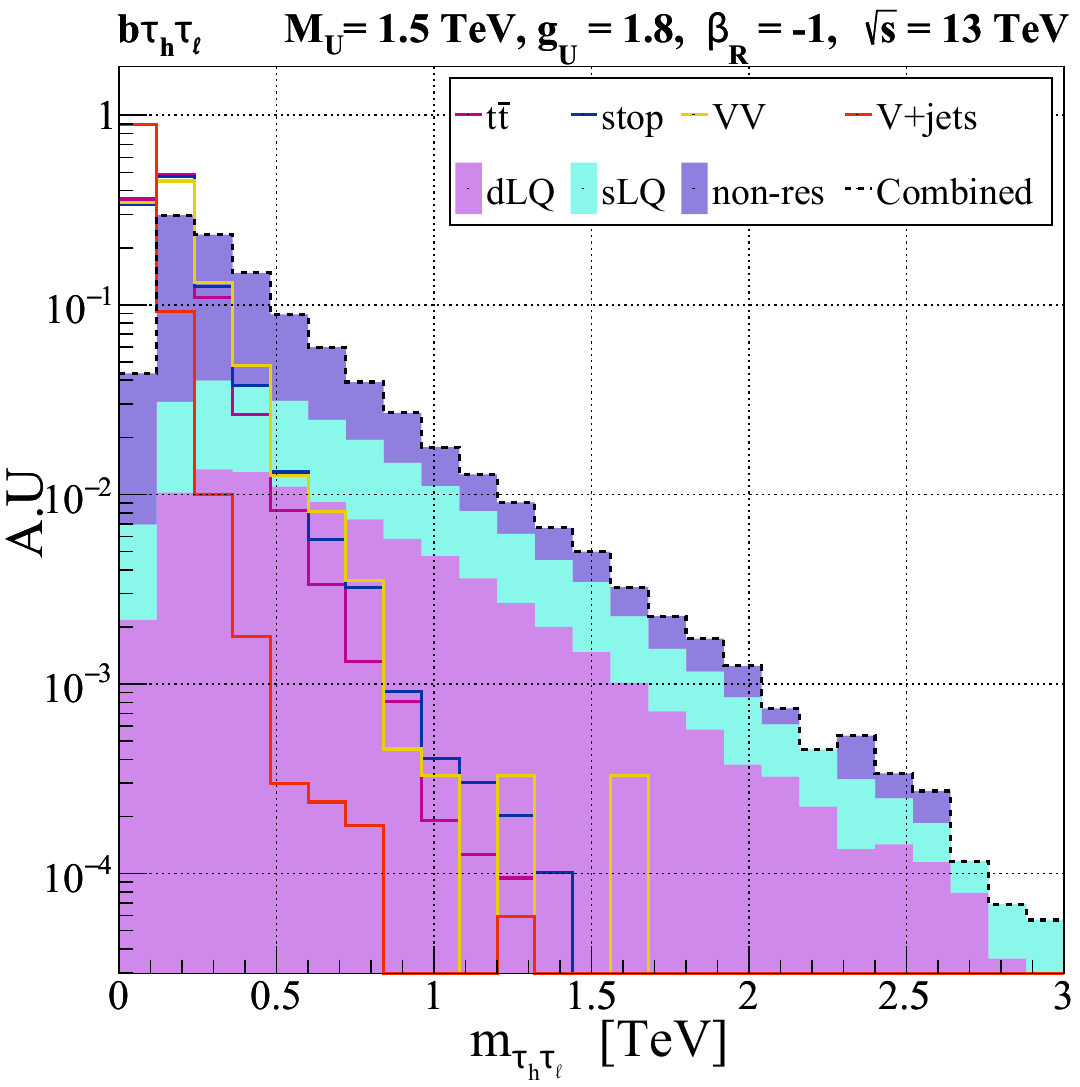}
    \includegraphics[height=4.9cm]{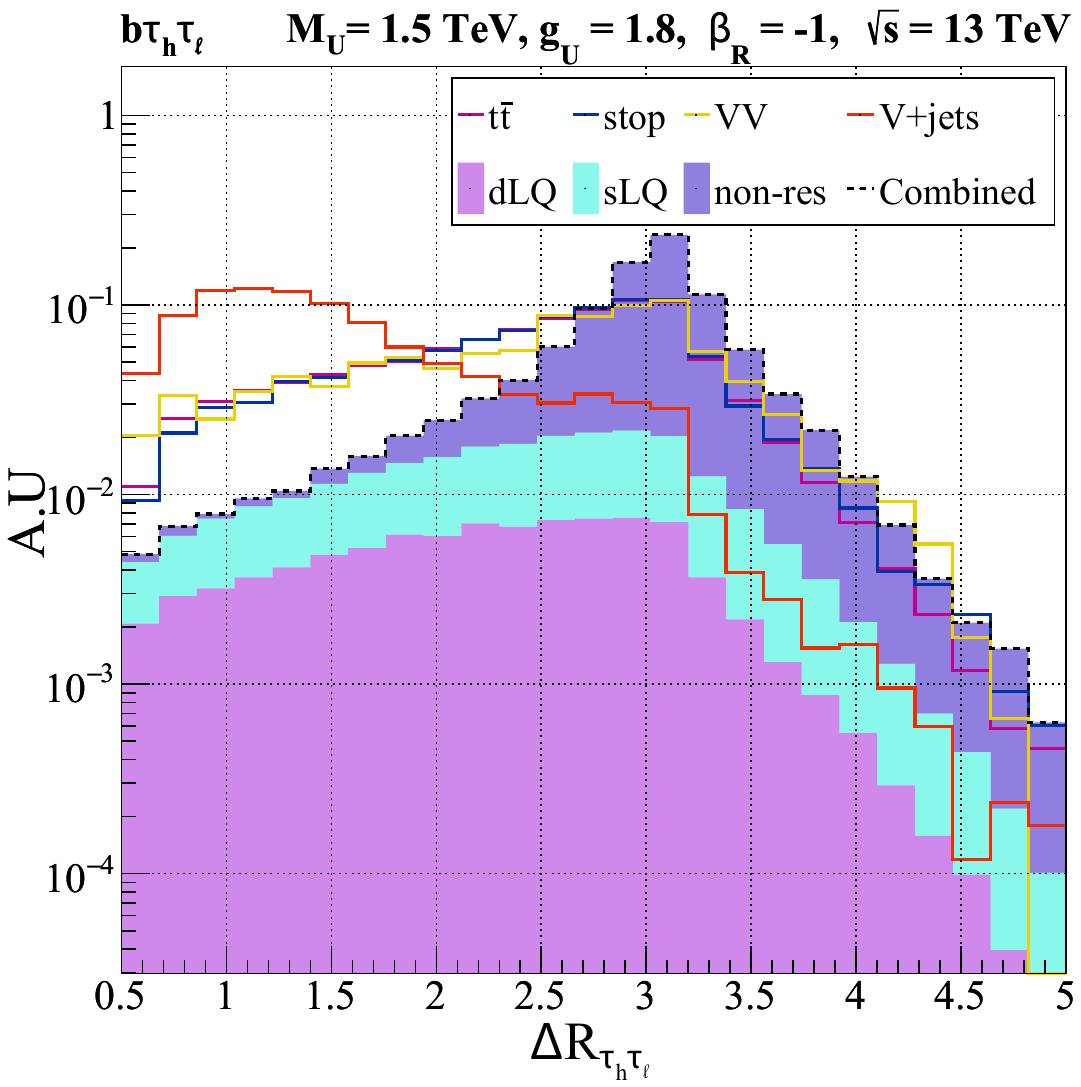}
    \includegraphics[height=4.9cm]{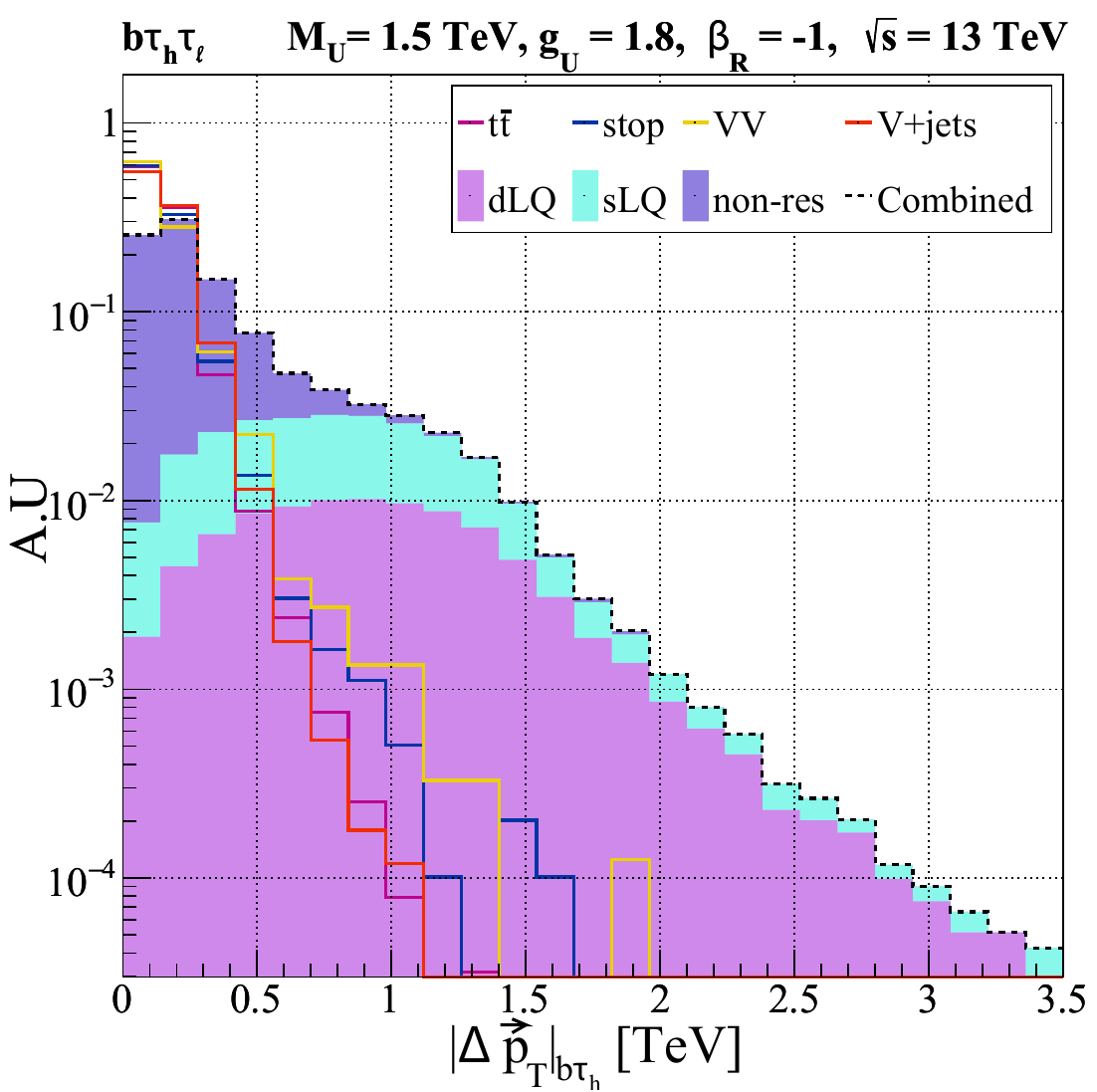}    
    \caption{$S_{\mathrm{T}}^{MET}$, $m_{\tau_{\mathrm h} \tau_{\ell}}$, $\Delta R_{\tau_{\mathrm h}\tau_{\ell}}$, $|\Delta \vec{p}_T|_{b \tau_{\mathrm h}}$ signal and background distributions for the $b\tau_h\tau_\ell$ channel. The signal distributions are generated for a benchmark sample with $\lq$ mass of $1.5 \tev$  maximally coupled to right-handed currents. The combined distribution (shown as a stacked histogram) is the sum of the distributions, correctly weighted according to their respective cross-sections, assuming a coupling $g_U = 1.8$.}
    \label{fig:sT(TeV)_wRHC}
\end{figure}

Figure~\ref{fig:sT(TeV)_wRHC} shows some relevant topological distributions, including $S_{\mathrm{T}}^{MET}$ on the top, for the $\bq\, \tau_{\mathrm h} \tau_{\ell}$ category.  
In the Figure we include all signal production modes to this channel, with each component weighted with respect to their total contribution to the combined signal. The combined signal distribution is normalised to unity. We also show all background processes contributing to this channel, each of them individually normalised to unity. We find that the combined signal is dominated by s$\lq$ production for large values of $S_{\mathrm{T}}^{MET}$, while non-res production dominates for small $S_{\mathrm{T}}^{MET}$. Interestingly, the backgrounds also sit at low $S_{\mathrm{T}}^{MET}$ values, since $S_{\mathrm{T}}^{MET}$ is driven by the mass scale of the SM particles being produced, in this case top quarks and Z/W bosons. This suggest that the s$\lq$ and d$\lq$ signals can indeed be separated from the SM background. As expected, the $S_{\mathrm{T}}^{MET}$ s$\lq$ and d$\lq$ signal distributions have a mean near $M_U$, representative of resonant production, and a broad width as expected for large mass $M_{U}$ hypotheses when information about the $z$-components of the momenta of objects is not utilised in the $S_{\mathrm{T}}^{MET}$ calculation.  

Figure~\ref{fig:sT(TeV)_wRHC} (second from the top) shows the reconstructed mass of the ditau system, for the $\bq \tau_{\textrm{h}}\tau_{\ell}$ search channel. Since the two $\tau$ candidates in signal events arise from different production vertices (e.g., each $\tau$ candidate in d$\lq$ production comes from a different $\lq$ decay chain), the ditau mass distribution for signal scales as $m_{\tau_{\textrm{h}}\tau_{\ell}} \sim \pt(\tau_{\textrm{h}}) + \pt(\tau_{\ell})$, and thus has a tail which depends on $M_{U}$ and sits above the expected SM spectrum. On the other hand, the SM $m_{\tau_{\textrm{h}}\tau_{\ell}}$ distributions sit near $m_{\textrm{Z/W}}$ since the $\tau$ candidates in SM events arise from $\zb/\wb$ decays.

Figure~\ref{fig:sT(TeV)_wRHC} (third from the top) shows the $\Delta R_{\tau_{\textrm{h}}\tau_{\ell}}$ distribution for the $\textrm{b}\tau_{\textrm{h}}\tau_{\ell}$ channel. In the case of the $\mathrm{p}\,\mathrm{p}\to\tau\tau$ non-res signal distribution, the two $\tau$ leptons must be back-to-back to preserve conservation of momentum. Therefore, the visible $\tau$ candidates, $\tau_{\textrm{h}}$ and $\tau_{\ell}$, give rise to a $\Delta R_{\tau_{\textrm{h}}\tau_{\ell}}$ distribution that peaks near $\pi$ radians. In the case of s$\lq$ production, although the $\lq$ and associated $\tau$ candidate must be back-to-back, the second $\tau$ candidate arising directly from the decay of the $\lq$ does not necessarily move along the direction of the $\lq$ (since the $\lq$ also decays to a b quark). As a result, the $\Delta R_{\tau_{\textrm{h}}\tau_{\ell}}$ distribution for the s$\lq$ signal process is smeared out, is broader, and has a mean below $\pi$ radians. On the other hand, the $\tau_{\textrm{h}}$ candidate in $\tq\bar{ \tq}$ events is often a jet being misidentified as a genuine $\tau_{\textrm{h}}$. When this occurs, the fake $\tau_{\textrm{h}}$ candidate can arise from the same top quark decay chain as the $\tau_{\ell}$ candidate, thus giving rise to small $\Delta R_{\tau_{\textrm{h}}\tau_{\ell}}$ values. This difference in the signal and background distributions provides a nice way for the ML algorithm to help decipher signal and background processes.

As noted above, the $|\Delta \vec{p}_{T}|$ distribution between the visible $\tau$ candidates and the b-quark jets is an important variable to help the BDT distinguish between signal and background processes. The discriminating power can be seen in Figure~\ref{fig:sT(TeV)_wRHC} (bottom), which shows the $|\Delta \vec{p}_{T}|$ between the $\tau_{\textrm{h}}$ and b-jet candidate of the $\textrm{b}\tau_{\textrm{h}}\tau_{\ell}$ channel. In the case of d$\lq$ production, the b quarks and $\tau$ leptons from the $\lq \to \textrm{b}\tau$ decay acquire transverse momentum of $p_{T} \sim M_{U}/2$. However, when the $\tau$ lepton decays hadronically (i.e. $\tau \to \tau_{\textrm{h}}\nu$), a large fraction of the momentum is lost to the neutrino. Therefore, the $|\Delta \vec{p}_{T}|_{\textrm{b}\tau_{\textrm{h}}}$ distribution for the d$\lq$ (and s$\lq$) process peaks below $M_{U}$/2. On the other hand, for a background process such as V+jets, the b jet arises due to initial state radiation, and thus must balance the momentum of the associated vector boson (i.e. $p_{T}(\textrm{b}) \sim p_{T}(\textrm{V}) \sim m_{\textrm{V}}$). Since the visible $\tau$ candidate is tyically produced from the V boson decay chain, its momentum (on average) is approximately $p_{T}(\tau_{\textrm{h}}) \sim p_{T}(\textrm{V})/4 \sim m_{\textrm{V}}/4$. Therefore, to first order, the $|\Delta \vec{p}_{T}|$ distribution for the V+jets background is expected to peak below the $m_{\textrm{V}}$ mass. 

\begin{figure}[]
    \centering
    \includegraphics[width=.95\linewidth]{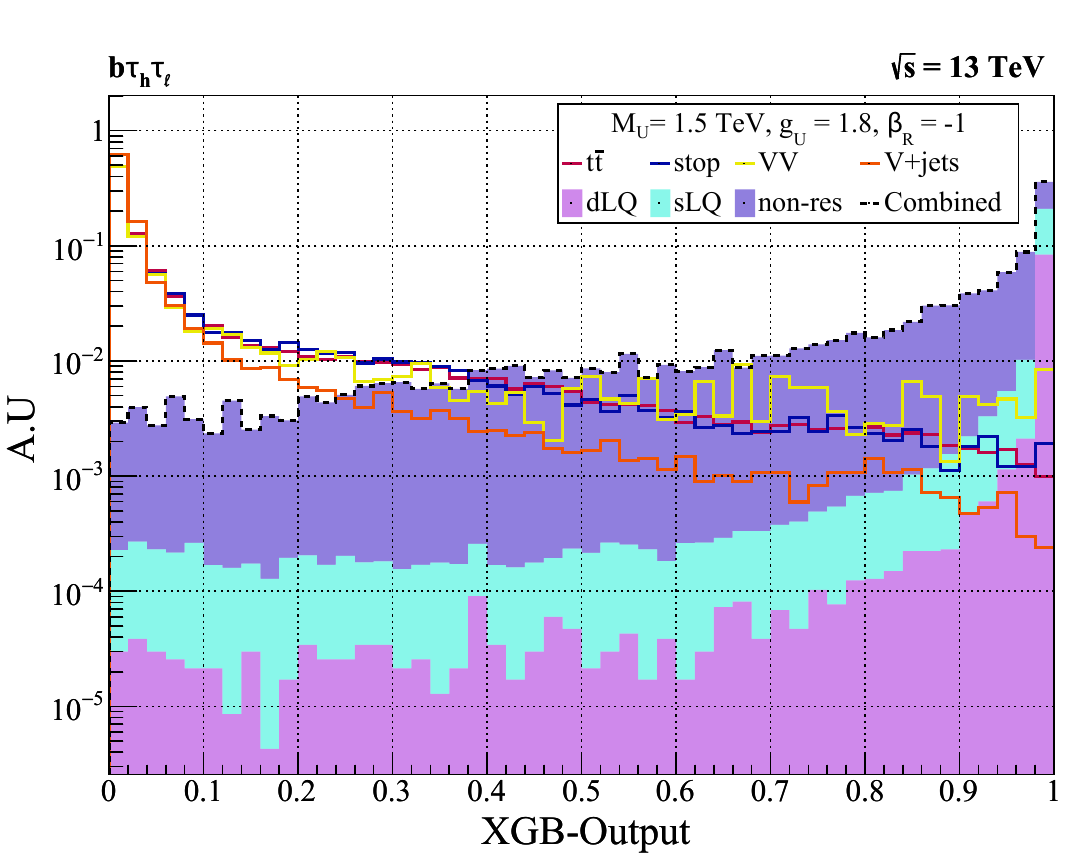}
    \caption{Postfit XGB-output normalised distribution in the $\bq\, \tau_{\mathrm h} \tau_{\ell}$ channel, for $\lq$ mass of 1.5\tev, constant coupling $g_U=1.8$, and maximally coupled to right-handed currents.}
    \label{fig:XGB_output}
\end{figure}
Lets us turn to the results of the $\bq\tau_{\mathrm h}\tau_\ell$ BDT classifier, which is shown in Figure~\ref{fig:XGB_output} for the different signal production modes and backgrounds. Similar to Figure~\ref{fig:sT(TeV)_wRHC}, the distribution for each individual signal production mode is weighted with respect to their total contribution to the combined signal. The background distributions and combined signal distribution are normalized to an area under the curve of unity. Figure~\ref{fig:XGB_output} shows the XGB distributions for a signal benchmark point with $M_{U} = 1.5$ TeV, $g_{U} = 1.8$, and $\beta_{R} = -1$. The XGB output is a value between 0 and 1, which quantifies the likelihood that an event is either signal-like (XGB output near 1) or background-like (XGB output near 0). We see that the presence of the s$\lq$ and d$\lq$ production modes is observed as an enhancement near a XGB output of unity, while the backgrounds dominate over the low end of the XGB output spectrum, especially near zero. In fact, over eighty percent of the sLQ and dLQ distributions reside in the last two bins, XGB output greater than 0.96, while more than sixty percent of the backgrounds fall in the first two bins, XGB output less than 0.04. %\JPP{23\% lie on the left of XGB = 0.7}.
It is also interesting to note that in comparison to the sLQ and dLQ distributions in Figure~\ref{fig:XGB_output}, non-res is broader and not as narrowly peaked near XGB output of 1, which is expected due to the differences in kinematics described above. Overall, if we focus on the last bin in this distribution, we find approximately 0.2\% of the background, in contrast to 22\% of the non-res, 78\% of the sLQ, and 91\% of the dLQ signal distributions. These numbers highlight the effectiveness of the XGB output in reducing the background in the region where the signal is expected.

The output signal and background distributions of the XGB classifier, normalised to their cross section times pre-selection efficiency times luminosity, are used to perform a profile binned likelihood statistical test in order to determine the expected signal significance. The estimation is performed using the \texttt{RooFit}~\cite{RooFit} package, following the same methodology as in Refs.~\cite{Barbosa:2022mmw, Florez:2021zoo, Florez:2019tqr, Florez:2018ojp, Florez:2017xhf, VBFZprimePaper, Florez:2016lwi, U1T3R, mSUGRApaper, SupercriticalString, ConnectingPPandCosmology, VBF1, DMmodels2, VBFSlepton, VBFStop, VBFSbottom}. The value of the significance ($Z_{sig}$) is measured using the probability to obtain the same outcome from the test statistic in the background-only hypothesis, with respect to the signal plus background hypothesis. This allows for the determination of the local p-value and thus the calculation of the signal significance, which corresponds to the point where the integral of a Gaussian distribution between $Z_{sig}$ and $\infty$ results in a value equal to the local p-value. 

Systematic uncertainties are incorporated as nuisance parameters, considering log-priors for normalization and Gaussian priors for shape uncertainties. Our consideration of systematic uncertainties includes both experimental and theoretical effects, focusing on the dominant sources of uncertainty. Following~\cite{lumiRef}, we consider a 3\% systematic uncertainty on the measurement of the integrated luminosity at the LHC. A 5\% uncertainty arises due to the choice of the parton distribution function used for the MC production, following the PDF4LHC prescription~\cite{Butterworth:2015oua}. The chosen PDF set only has an effect on the overall expected signal and background yields, but the effect on the shape of the XGB output distribution is negligible. Reference~\cite{CMS_DeepTau} reports a systematic uncertainty of 2-5\%, depending on the $p_{\textrm{T}}$ and $\eta$ of the $\tau_{\textrm{h}}$ candidate. Therefore, we utilize a conservative 5\% uncertainty per $\tau_{\textrm{h}}$ candidate, independent of $p_{\textrm{T}}$ and $\eta$, which is correlated between signal and background processes with genuine $\tau_{\textrm{h}}$ candidates, and correlated across XGB bins for each process. We assumed a 5\% $\tau_{\textrm{h}}$ energy scale uncertainty, independent of $p_{\textrm{T}}$ and $\eta$, following the CMS measurements described in~\cite{CMS_DeepTau}.  Finally, we assume a conservative 3\% uncertainty per b-jet candidate, following reference~\cite{CMSbtag}, and an additional 10\% uncertainty in all the background predictions to account for possible mismodeling by the simulated samples. The uncertainties on the background estimates are typically derived from collision data using dedicated control samples that have negligible signal contamination and are enriched with events from the specific targeted background. The systematic uncertainties on the background estimates are treated as uncorrelated between background processes.

\begin{figure}[]
    \centering
        \begin{subfigure}[b]{\linewidth}
            \includegraphics[height=5.8cm]{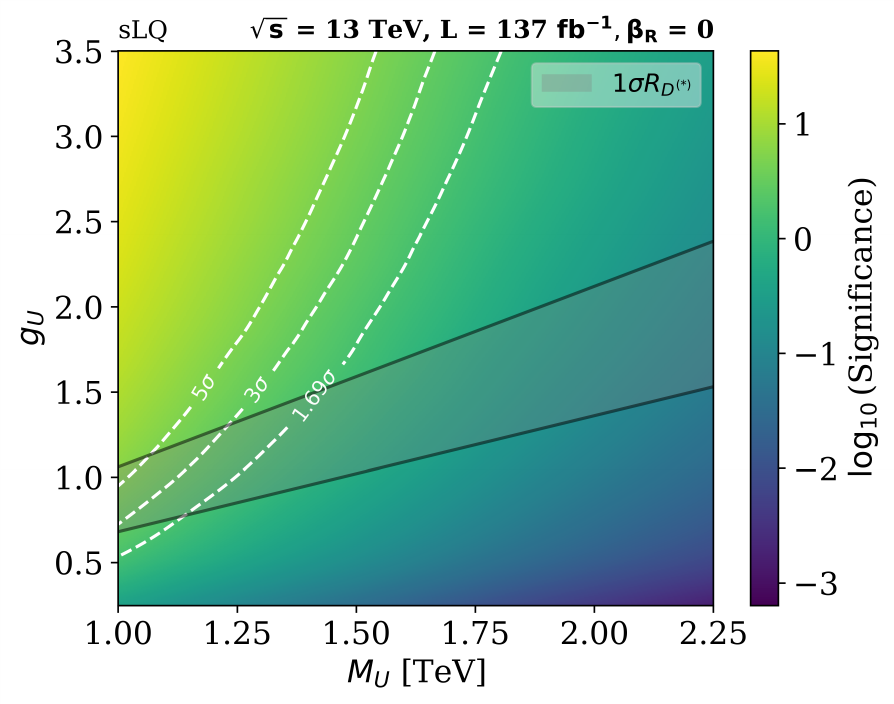}
            \hspace{4pt}
        \end{subfigure}
        \begin{subfigure}[b]{\linewidth}
            \includegraphics[height=5.8cm]{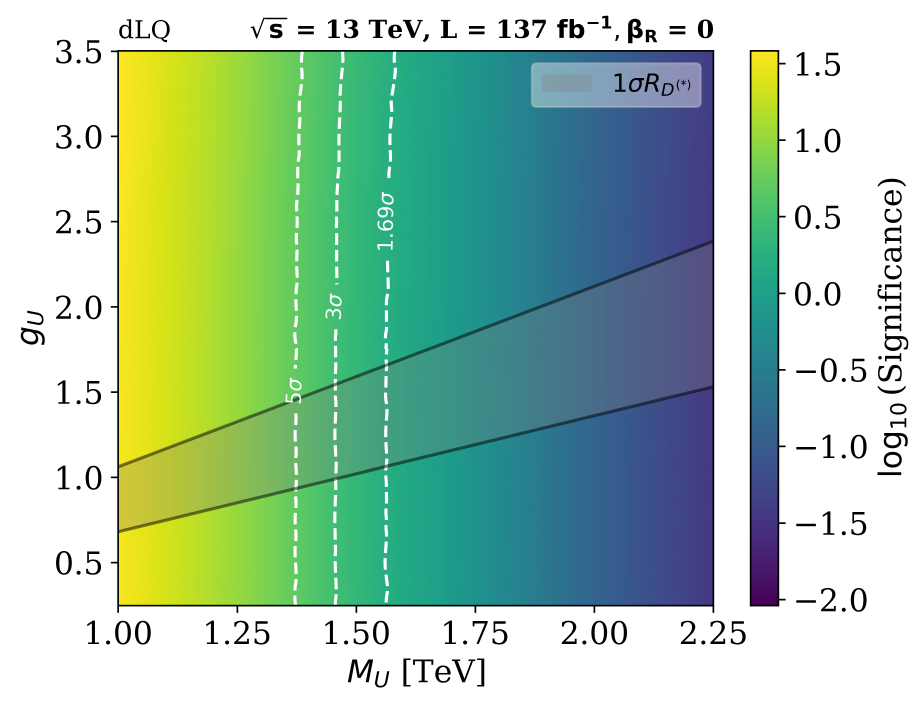}
        \end{subfigure}     
        \begin{subfigure}[b]{\linewidth}
            \includegraphics[height=5.8cm]{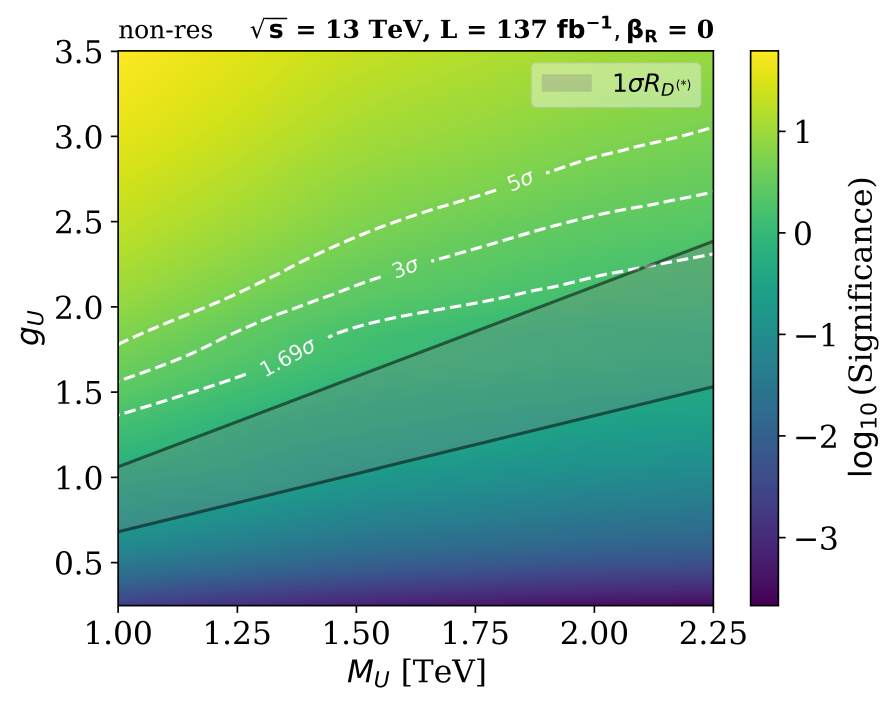}
        \end{subfigure}    
 
    \caption{Signal significance for different coupling scenarios and $\lq$ masses, without right-handed currents, using the combination of all search channels. The results pertaining to  s$\lq$, d$\lq$ and non-res production are displayed respectively from the top.  These results are for $\sqrt{s} = 13 \tev$ and $137 \fb^{-1}$.}
    \label{fig:heatmapssignificance}
\end{figure}

\section{Results}
\label{sec:results}

The expected signal significance for s$\lq$, d$\lq$ and non-res production, and their combination, is presented in Figure~\ref{fig:heatmapssignificance}. Here, the significance is shown as a heat map in a two dimensional plane of $g_U$ versus $M_U$, considering exclusive couplings to left-handed currents, \textit{i.e.} ${\rm BR}(\lq \to \bq\,\tau)=\tfrac12$. The dashed lines show the contours of constant signal significance. The $1.69 \sigma$ contour represents exclusion at 95\% confidence level, and the 3-5$\sigma$ contours represent potential discovery. The grey band defines the set of $\set{M_{U},g_{U}}$ values that can explain the $\Bm$-meson anomalies, $C_U\sim 0.01$ for this scenario. The estimates are performed under the conditions for the second run, RUN-II, of the LHC ($\sqrt{s} = 13 \tev$ and $L = 137 \fb^{-1}$). We find that the d$\lq$ interpretation plot (Figure ~\ref{fig:heatmapssignificance} second from the top) does not depend on $g_{U}$, which is expected due to d$\lq$ production arising exclusively from interactions with gluons. For this reason, the d$\lq$ production process provides the best mode for discovery when $g_{U}$ is small. On the other hand, the non-res channel is more sensitive to changes in the coupling parameter $g_U$, as its production cross-section depends on $g_{U}^{4}$. Therefore, the non-res production process provides the best mode for discovery when $g_{U}$ is large. These results confirm the expectations from previous analyses (see for instance~\cite{Schmaltz:2018nls}), in the sense that the d$\lq$ and non-res processes complement each other nicely at low and high $g_{U}$ scenarios. The s$\lq$ channel combines features from both the d$\lq$ and non-res channels, in principle making it an interesting option to explore different scenarios and gain a better understanding of $\lq$ properties, but the evolution of the signal significance in the full phase space is more complicated as it involves resonant $\lq$ production with a cross-section that depends non-trivially on $M_{U}$, $g_{U}$, and the $\lq$ coupling to gluons. However, Figure~\ref{fig:heatmapssignificance} shows that the s$\lq$ production process can provide complementary and competitive sensitivity to the non-res and d$\lq$ processes, in certain parts of the phase space.

\begin{figure}[t]
    \centering
       \begin{subfigure}[b]{\linewidth}
            \includegraphics[height=5.8cm]{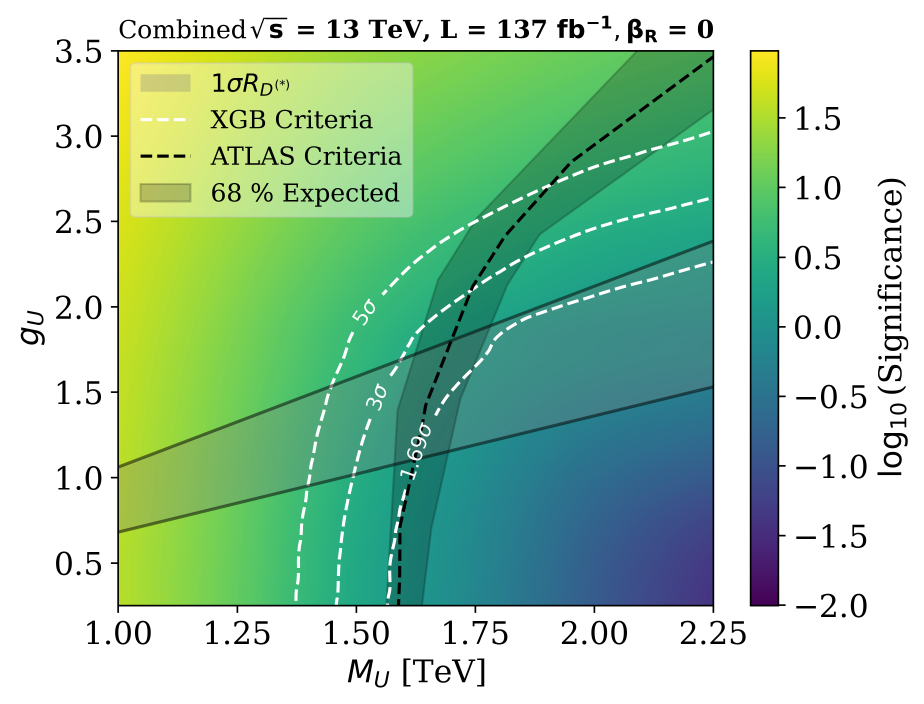}
        \end{subfigure}    
       \begin{subfigure}[b]{\linewidth}
            \includegraphics[height=5.8cm]{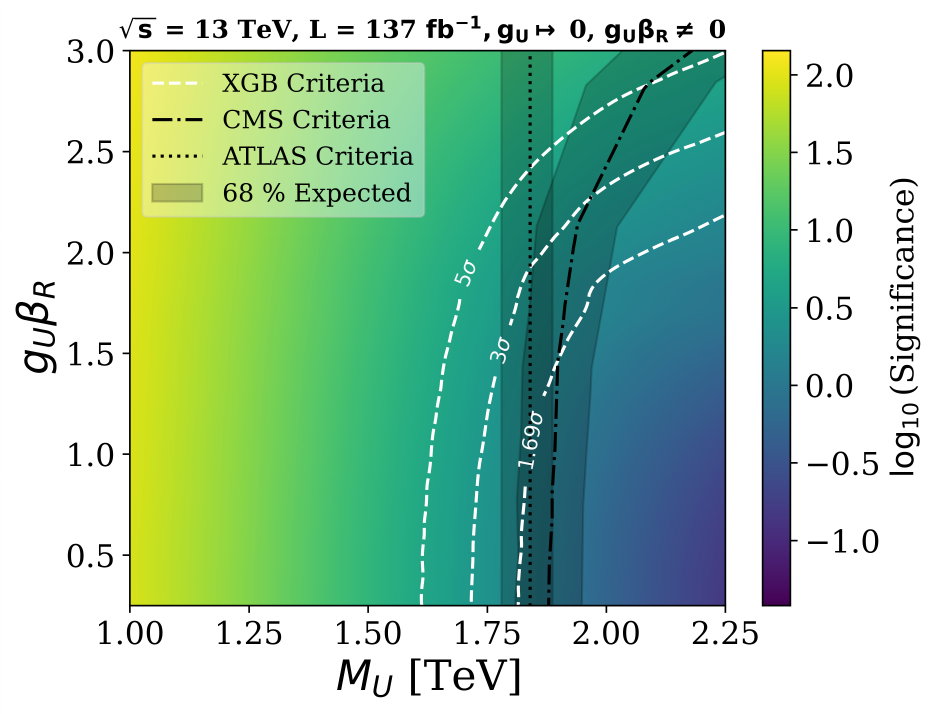}
        \end{subfigure}    
    \caption{The top (bottom) panel shows signal significance comparison with ATLAS~\cite{ATLAS_7A} (CMS and ATLAS~\cite{ LQS_CMS_2022_results_comparison, ATLAS_Vertical_Line}) background only hypothesis, for the combination of all channels, with uniquely coupling to left-handed (right-handed) currents. The estimates are performed at $\sqrt{s} = 13 \tev$ and $137 \fb^{-1}$.}
    \label{fig:heatmapscomparingcms}
    %It also includes a comparison with CMS results (report CMS-PAS-EXO-19-016)
\end{figure}

The top panel of Figure~\ref{fig:heatmapscomparingcms} presents the sensitivity of all signal production processes combined, and compares our expected exclusion region with the latest one from the ATLAS Collaboration~\cite{ATLAS_7A}. The comparison suggests that our proposed analysis strategy provides better sensitivity than current methods being carried out at ATLAS, especially at large values of $g_U$. In particular, we find that with the $\textrm{pp}$ data already available from RUN-II, our expected exclusion curves begin to probe solutions to the B-anomalies for $\lq$ masses up to $2.25\tev$.

Figure~\ref{fig:heatmapscomparingcms} shows the expected signal significance considering ${\rm BR}(\lq \to \bq\,\tau) = 1$, in order to compare our analysis with the corresponding results from the CMS~\cite{LQS_CMS_2022_results_comparison} and ATLAS~\cite{ATLAS_Vertical_Line} Collaborations. Let us emphasize again that ${\rm BR}(\lq \to \bq\,\tau)$ depends on $\beta_R$, as illustrated on the top panel of Figure~\ref{fig:branching_ratios}. Thus, although the ${\rm BR}(\lq \to \bq\,\tau) = 1$ scenario is a possible physical case, it does not solve the observed anomalies in the $R_{D^{(*)}}$ ratios, as it corresponds to the case where LQs couple exclusively to right-handed currents.

With this in mind, the scenario studied by CMS in~\cite{LQS_CMS_2022_results_comparison} considers couplings only to left-handed currents, setting artificially the condition ${\rm BR}(\lq \to \bq\,\tau) = 1$. In order to compare, we scale the efficiency$\times$acceptance of our selection criteria for $\beta_R=0$, by a factor of 2.0 for s$\lq$ and 4.0 for d$\lq$. According to Figure~\ref{fig:heatmapscomparingcms}, the ML approach that we have followed again suggests an optimisation of the signal and background separation, having the potential of improving the regions of exclusion (1.69 $\sigma$) with respect to that of CMS. In the bottom panel of the Figure we have also included a similar exclusion by ATLAS~\cite{ATLAS_Vertical_Line}. However, since ATLAS only considers d$\lq$ production in the analysis, the results are not entirely comparable, so are included only as a reference. 

\begin{figure}[t]
    \centering
    \begin{subfigure}[b]{0.495\textwidth}
            \includegraphics[height=6.3cm]{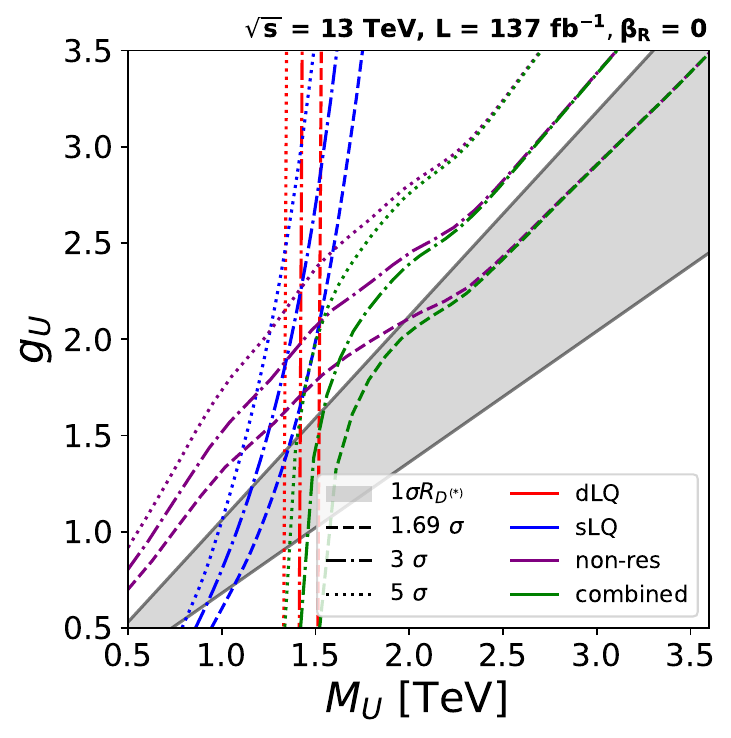}
    \end{subfigure}
    \begin{subfigure}[b]{0.495\textwidth}
            \includegraphics[height=6.3cm]{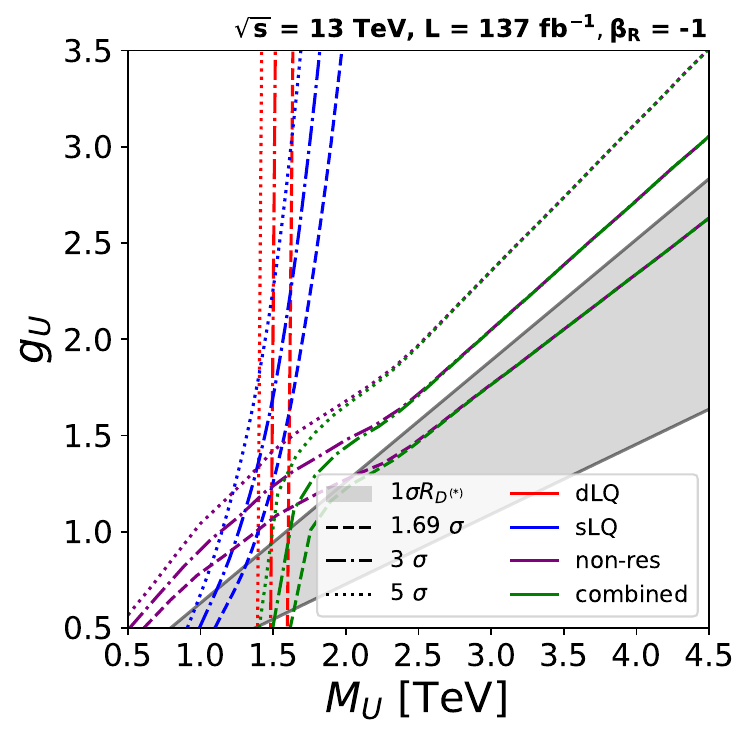}
    \end{subfigure}
    \caption{Signal significance for different coupling scenarios and $\lq$ masses for all channels. This plot summarizes our results with $\beta_{R} = 0$ (without right-handed currents) and $\beta_{R} = -1$ (maximally coupled to right-handed currents). The estimates are performed at $\sqrt{s} = 13 \tev$ and $137 \fb^{-1}$.}
    \label{fig:significance137ifb}
\end{figure}
We now turn to the role of $\beta_R$, and our capacity of probing the regions solving the B-meson anomalies. Figure~\ref{fig:significance137ifb} shows the maximum significant contours, under LHC RUN-II conditions, for the different $\lq$ production mechanisms and their combination, considering scenarios with only left-handed currents ($\beta_R=0$, top) and with maximal right-handed currents ($\beta_R=-1$, bottom). We find a noticeable improvement in signal significance in all channels when taking $\beta_R=-1$, as is expected from the increase in ${\rm BR}(\lq \to \bq\,\tau)$ branching ratio and production cross-sections (see Figure~\ref{fig:branching_ratios}). However, the region solving the B-meson anomalies also changes, preferring lower values of $g_U$, such that in both cases we find ourselves just starting to probe this region at large $M_U$.

\begin{figure}[]
    \centering
    \includegraphics[height=6.5cm]{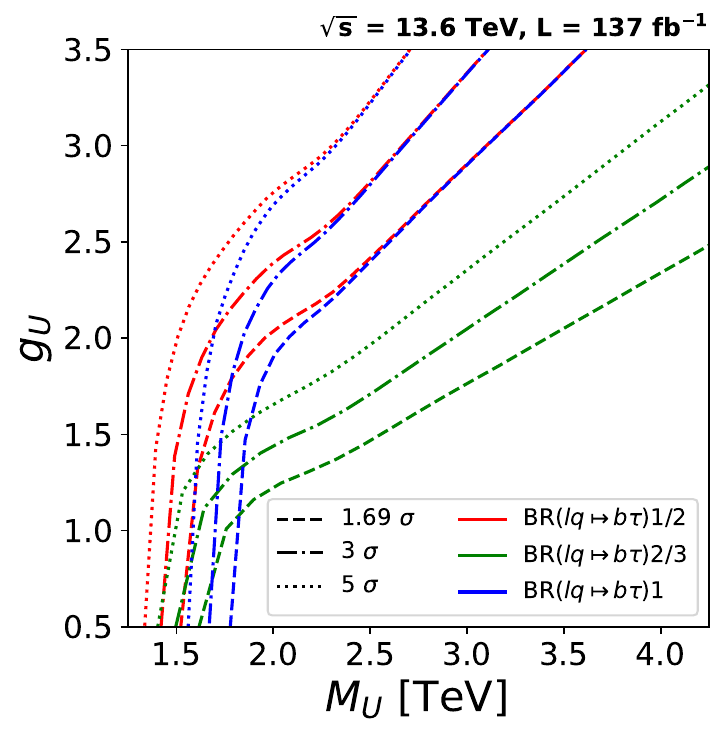}
    \caption{Signal significance for different coupling scenarios and $\lq$ masses, 
    considering the case without coupling to right-handed currents ${\rm BR}(\lq \to \bq\,\tau) = \tfrac{1}{2}$, the case maximally coupled to right- and left-handed currents ${\rm BR}(\lq\to b\,\tau) = \tfrac23$, and the case uniquely coupled to right-handed currents ${\rm BR}(\lq \to \bq\,\tau) = 1$. The estimates are performed at $\sqrt{s} = 13 \tev$ and $137 \fb^{-1}$.}
    \label{fig:combinedsigniBRs}
\end{figure}

The combined significance contours for the different ${\rm BR}$ scenarios that have been considered is presented in Figure~\ref{fig:combinedsigniBRs}. These contours illustrate the regions of exclusion for the three cases of interest, namely exclusive left-handed currents (${\rm BR}(\lq \to \bq\,\tau) = \tfrac{1}{2}$, $\beta_R=0$), maximal left and right couplings (${\rm BR}(\lq\to b\,\tau) = \tfrac23$, $\beta_R=-1)$, and exclusive right-handed currents (${\rm BR}(\lq \to \bq\,\tau) = 1$, $g_U\to0,\,g_U\beta_R=1$). For small $g_U$, we find that the exclusive right-handed scenario is most sensitive, while the exclusive left-handed case is the worst. The reason for this is that this region is excluded principally by d$\lq$ production, such that having the largest branching ratio is crucial in order to have a large number of events. For larger couplings, both exclusive scenarios end up having similar exclusion regions, with the $\beta_R=-1$ case being significantly more sensitive. The reason in this case is that the exclusion is dominated by non-res, which has a much larger production cross-section if both currents are turned on.

\begin{figure}[]
    \centering
    \begin{subfigure}[b]{0.495\textwidth}
        \includegraphics[height=6.5cm,]{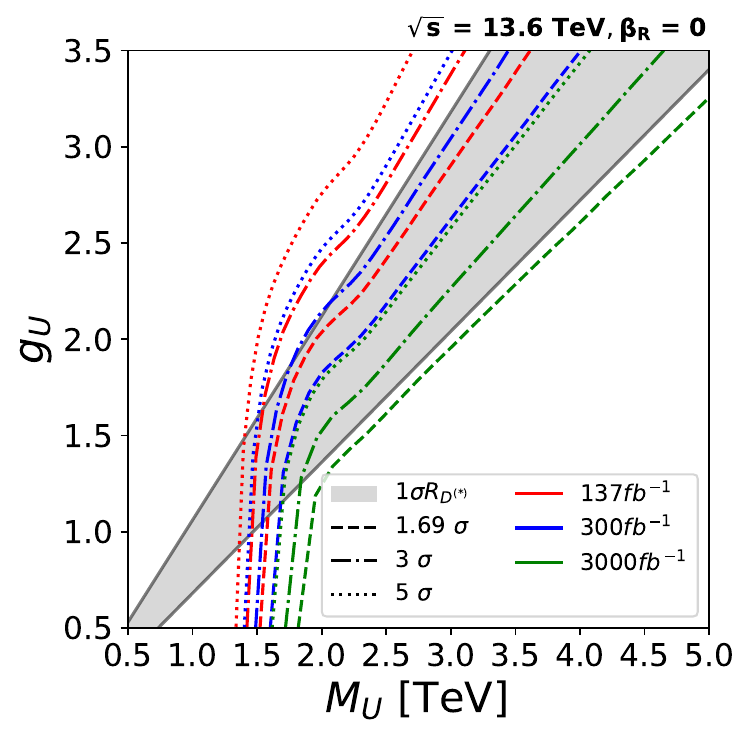}
    \end{subfigure}
    \begin{subfigure}[b]{0.48\textwidth}
        \includegraphics[height=6.5cm]{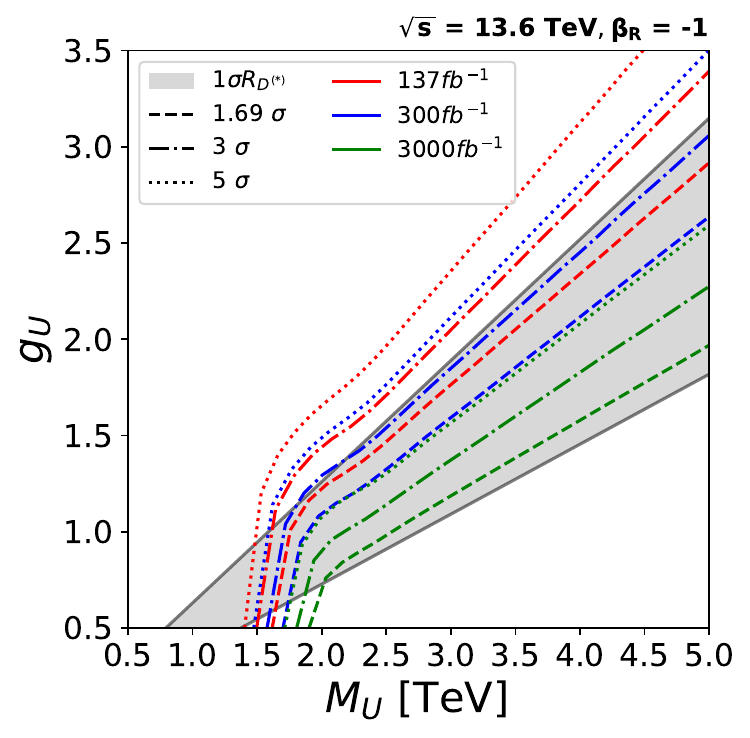}
    \end{subfigure}
    \caption{Projected signal significance for different coupling scenarios and $\lq$ masses maximally coupled to right-handed currents. The estimates are performed at $\sqrt{s} = 13.6 \tev$, $137 \fb^{-1}$, $300 \fb^{-1}$ and $3000 \fb^{-1}$.}
    \label{fig:combinedsigniLumis}
\end{figure}

In order to finalise our analysis of the LQ-only model, we show in Figure~\ref{fig:combinedsigniLumis} the expected combined significance in the relatively near future. For this, considering $\sqrt{s} = 13.6 \tev$, we show contours for the sensitivity corresponding to integrated luminosities of $137 \fb^{-1}$,  $300 \fb^{-1}$, and $3000 \fb^{-1}$, for scenarios with only left-handed currents (top) and with maximal coupling to right-handed currents (bottom). Note that for $\beta_R = 0$ ($\beta_R = -1$), couplings $g_U$ close to 3.18 (1.85)  and $M_U = 5.0 \tev$ can be excluded with $1.69 \sigma$ significance for the high luminosity LHC era, allowing us to probe the practically the entirety of the B-meson anomaly favored region. Note that the background yields for the high luminosity LHC might be larger due to pileup effects. Nevertheless, as it was mentioned in Section~\ref{sec:strategyandsimulation}, we have included a conservative 10\% systematic uncertainty associated with possible fluctuations on the background estimations. Although effects from larger pileup might be significant, they can be mitigated by improvements in the algorithms for particle reconstruction and identification, and also on the data-analysis techniques.

As commented on the Introduction, non-res production can be significantly affected by the presence of a companion $\zb'$, which provides additional s-channel diagrams that add to the total cross-section and can interfere destructively with the $\lq$ t-channel process (see Figures~\ref{fig:xsinterference}
and~\ref{fig:interference}). From our previous results, we see that non-res always is of high importance in determining the exclusion region, particularly at large $M_U$ and $g_U$, meaning it is crucial to understand how this role is affected in front of a $\zb'$ with similar mass.

\begin{figure}[]
\centering
    \begin{subfigure}[b]{.94\linewidth}
    \includegraphics[height=6.1cm]{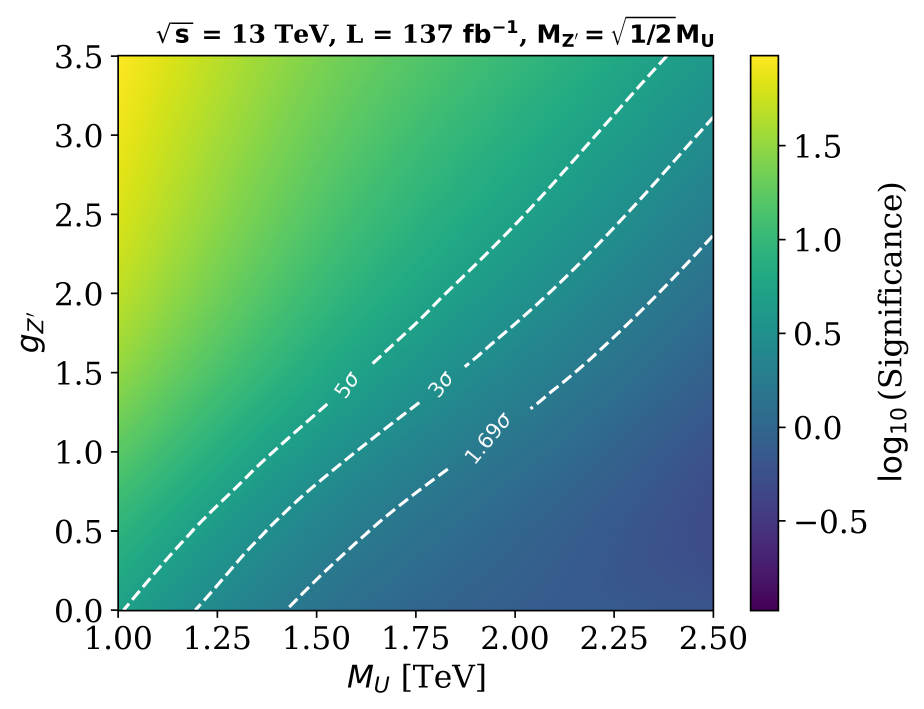}
    \end{subfigure}
    \begin{subfigure}[b]{.94\linewidth}
    \includegraphics[height=6.1cm]{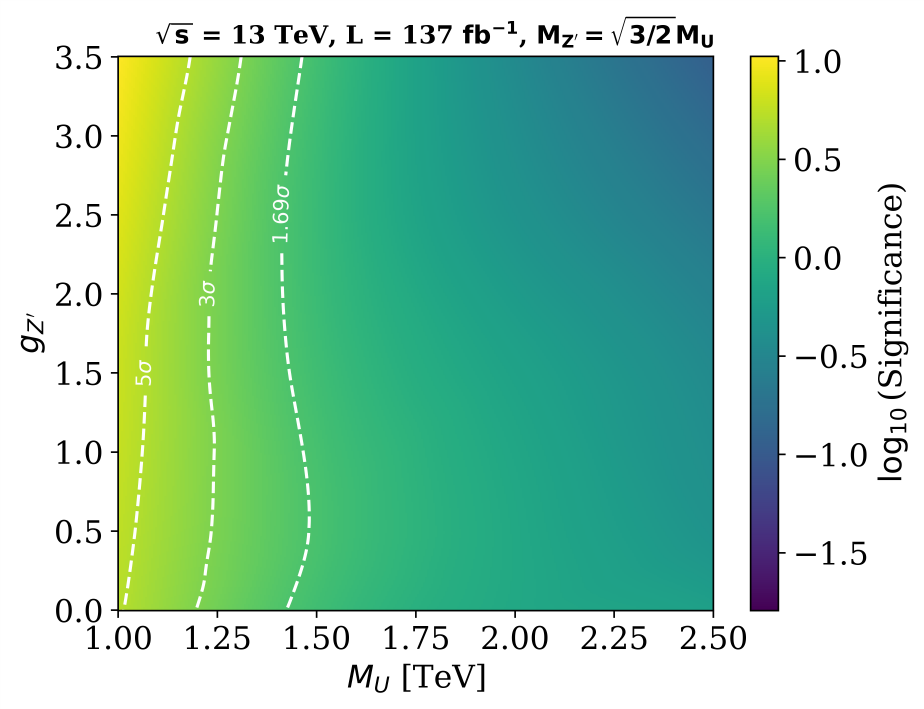}
    \end{subfigure}
    \caption{Change on the non-res signal significance for different $Z^{\prime}$ coupling scenarios and $\lq$ masses. The estimates are performed at $\sqrt s=13.0 $\tev, $\beta_R=0$, $g_U = 1.8$, $M_{\zb^{\prime}} = \sqrt{1/2} M_{U}$ (top), and $M_{\zb^{\prime}} = \sqrt{3/2} M_{U}$ (bottom).}
\label{fig:sensitivity_change}
\end{figure}

The change in sensitivity on the non-res signal significance due this interference effect with the $\zb^{\prime}$ boson is shown in Figure~\ref{fig:sensitivity_change}. We consider two opposite cases for the $\zb'$ mass: $M^2_{\zb'} = M^2_U/2$ (top) and $M^2_{\zb'} = 3\,M^2_U/2$ (bottom). Our results are shown on the $g_{\zb'}$ - $M_U$ plane, for a fixed $g_U=1.8$ and $\beta_R=0$. For the $M^2_{\zb'} = M^2_U/2$ scenario, there is an overall increase in the total cross-section, with a larger $g_{\zb'}$ implying a larger sensitivity. This means that our ability to probe smaller values of $g_U$ could be enhanced, as a given observation would be reproduced with both a specific $g_U$ and vanishing $g_{\zb'}$, or a smaller $g_U$ with large $g_{\zb'}$. Thus, for a large enough $g_{\zb'}$, it could be possible to enhance non-res to the point that the entire region favoured by $\Bm$-anomalies could be ruled out. In contrast, for $M^2_{\zb'} = 3\,M^2_U/2$ the cross-section is strongly affected by the large destructive interference, such that a larger $g_{\zb'}$ does not necessarily imply an increase in sensitivity. In fact, as can be seen in the bottom panel, for large $M_U$ the significance is reduced as $g_{\zb'}$ increases, leading to the opposite conclusion than above, namely, that a large $g_{\zb'}$ could reduce the effectiveness of non-res.

\begin{figure}[]
\centering
    \begin{subfigure}[b]{.94\linewidth}
    \includegraphics[height=6.1cm, width=6cm]{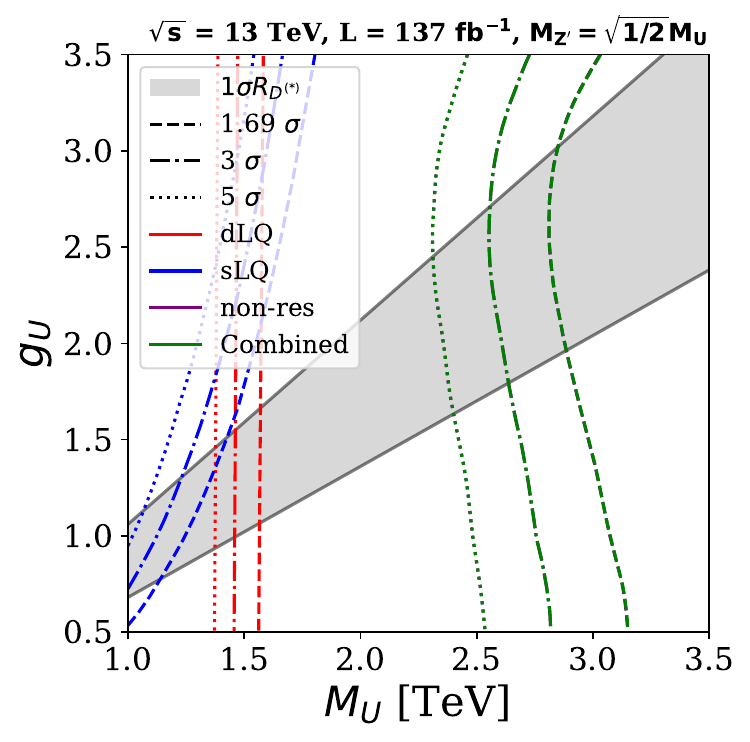}
    \end{subfigure}
    \begin{subfigure}[b]{.94\linewidth}
    \includegraphics[height=6.1cm, width=6cm]{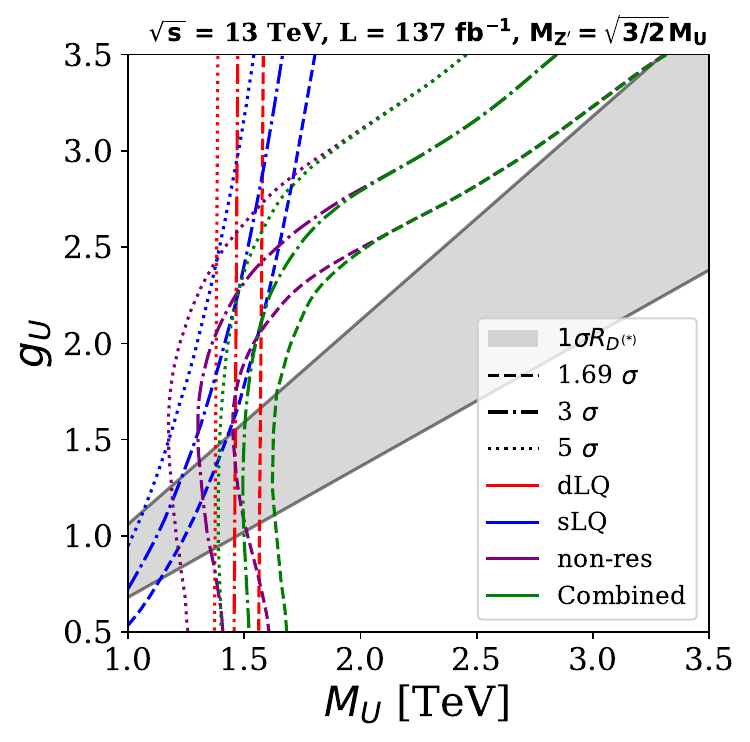}
    \end{subfigure}
    \caption{Signal significance for different coupling scenarios and $\lq$ masses, for all channels, with an additional $\zb'$ contribution to non-res production. We set $\beta_{R} = 0$ and $g_{\zb'}=3.5$, taking $M^2_{\zb'}$ equal to $M_U^2/2$  ($3M_U^2/2$) on the top (bottom) panel.}
\label{fig:sensitivity_gzp_fixed}
\end{figure}

The impact of the above can be seen in Figure~\ref{fig:sensitivity_gzp_fixed}, which shows our previous sensitivity curves on the $M_U-g_U$ plane, but this time with a $\zb'$ contribution to non-res. We use the same values of $M_{\zb'}$ as before, but fix $g_{\zb'}=3.5$. For smaller $M_{\zb'}$ (top), the non-res contribution is enhanced so much, that both s$\lq$ and d$\lq$ play no role whatsoever in determining the exclusion region. We find that, for small $g_U$, the sensitivity is dominated by $\zb'$ production such that, since $M_U$ is related to $M_{\zb'}$, $\lq$ masses up to $\sim3\tev$ are excluded. This bound is slightly relaxed for larger values of $g_U$, which is attributed to destructive interference effects due to an increased $\lq$~contribution.

The bottom panel of Figure~\ref{fig:sensitivity_gzp_fixed} shows that case where $M_{\zb'}$ is larger than $M_U$. As expected from our previous discussion, the behaviour and impact of non-res is modified. For small $g_U$, we again have the pure $\zb'$ production dominating the non-res cross-section, leading to a null sensitivity on $g_U$, similar to what happens in dLQ. In contrast, for very large $g_U$, we find that the pure $\lq$ non-res production is the one that dominates, and we recover sensitivity regions with a slope similar to those shown in Figures~\ref{fig:heatmapssignificance}-\ref{fig:combinedsigniLumis}, shifted towards larger values of $g_U$. For intermediate values of this coupling, the destructive interference have an important effect again, twisting the exclusion region slightly towards the left. Still, even in this case, we find that s$\lq$ plays a marginal role in defining the combined exclusion region, and that the final result again depends primarily on d$\lq$ and non-res production.

\section{Discussion and conclusions}
\label{sec:discusion}

Experimental searches for $\lq$s with preferential couplings to third generation fermions are currently of great interest due to their potential to explain observed tensions in the $R(D)$ and $R(D^{*})$ decay ratios of $\Bm$ mesons with respect to the SM predictions. Although the LHC has a broad physics program on searches for $\lq$s, it is very important to consider the impact of each search within wide range of different theoretical assumptions within a specific model. In addition, in order to improve the sensitivity to detect possible signs of physics beyond the SM, it is also important to strongly consider new computational techniques based on machine learning (ML). Therefore, we have studied the production of $U_1$ $\lq$s with preferential couplings to third generation fermions, considering different couplings, masses and chiral currents. These studies have been performed considering $\mathrm{p}\,\mathrm{p}$ collisions at $\sqrt{s} = 13\tev$ and $13.6\tev$ and different luminosity scenarios, including projections for the high luminosity LHC. A ML algorithm based on boosted decision trees is used to maximize the signal significance. The signal to background discrimination output of the algorithm is taken as input to perform a profile binned-likelihood test statistic to extract the expected signal significance. 

The expected signal significance for s$\lq$, d$\lq$ and non-res production, and their combination, is presented as contours on a two dimensional plane of $g_U$ versus $M_U$. We present results for the case of exclusive couplings to left-handed, mixed, and exclusive right-handed currents. For the first two, the region of the phase space that could explain the $\Bm$ meson anomalies is also presented. We confirm the findings of previous works that the largest production cross-section and best overall significance comes from the combination of d$\lq$ and non-res production channels. We also find that the sensitivity to probe the parameter space of the model is highly dependent on the chirality of the couplings. Nevertheless, the region solving the $\Bm$-meson anomalies also changes with each choice, such that in all evaluated cases we find ourselves just starting to probe this region at large $M_U$.

Our studies compare our exclusion regions with respect to the latest reported results from the ATLAS and CMS Collaborations. The comparison suggests that our ML approach has a better sensitivity than the standard cut-based analyses, especially at large values of $g_U$. In addition, our projections for the HL-LHC cover the whole region solving the B-anomalies, for masses up to $5.00\tev$.

Finally, we consider the effects of a companion $\zb^{\prime}$ boson on non-res production. We find that such a contribution can have a considerable impact on the LQ sensitivity regions, depending on the specific masses and couplings. In spite of this, we still consider non-res production as an essential channel for probing LQs in the future.

$$ $$

\begin{acknowledgments}
The authors would like to thank Gino Isidori for frutiful discussions. 
A.F, J.P, and C.R thank the constant and enduring financial support received for this project from the faculty of science at Universidad de Los Andes (Bogot\'a, Colombia) and the Colombian Science Ministry \-Min\-Cien\-cias, with the grant program 70141, contract number 164-2021. 
J.J.P.\ acknowledges funding by the {\it Direcci\'on de Gesti\'on de la Investigaci\'on} at PUCP, through grant No.\ DGI-2021-C-0020.
A.G acknowledges the funding received from the Physics \& Astronomy department at Vanderbilt University and the US National Science Foundation. This work is supported in part by NSF Award PHY-1945366 and a Vanderbilt Seeding Success Grant.
\end{acknowledgments}

%\appendix

%\newpage
% \bibliographystyle{utphys}
\bibliographystyle{unsrt}
\bibliography{apssamp}% Produces the bibliography via BibTeX.

\end{document}